\documentclass[
aps,
nofootinbib,
prd,
superscriptaddress,
tightenlines,
notitlepage,
twocolumn,
showpacs,
floatfix
]{revtex4-1}

\usepackage[colorlinks]{hyperref}
\usepackage{tabularx}
\usepackage{graphicx}
\usepackage{amssymb,amsmath,bm,tensor,braket,mathtools}
\usepackage{xcolor}
\usepackage[varg]{txfonts}
\usepackage{enumerate}
\usepackage{mathtools}
\usepackage[capitalize]{cleveref}
\usepackage[utf8]{inputenc}
\usepackage[normalem]{ulem}

\def\al{\alpha}
\def\be{\beta}
\def\ga{\gamma}
\def\de{\delta}
\def\ep{\epsilon}

\def\th{\theta}

\def\ka{\kappa}
\def\la{\lambda}

\def\si{\sigma}

\def\ph{\phi}

\def\Ga{\Gamma}

\def\Ph{\Phi}
\def\Ps{\Psi}

\def\mn{{\mu\nu}}
\def\abgd{{\al\be\ga\de}}

\def\lsim{\mathrel{\rlap{\lower4pt\hbox{\hskip1pt$\sim$}}
    \raise1pt\hbox{$<$}}}
\def\gsim{\mathrel{\rlap{\lower4pt\hbox{\hskip1pt$\sim$}}
    \raise1pt\hbox{$>$}}}
\def\sqr#1#2{{\vcenter{\vbox{\hrule height.#2pt
         \hbox{\vrule width.#2pt height#1pt \kern#1pt
         \vrule width.#2pt}
         \hrule height.#2pt}}}}

\def\prt{\partial}
\def\lrpartial{\raise 1pt\hbox{$\stackrel\leftrightarrow\partial$}}

\def\part2{\partial_\alpha \partial^\alpha}

\def\pt#1{\phantom{#1}}

\def\xx'{|\vec x -\vec x'|}

\def\b2{b^\al b_\al}

\newcommand{\beq}{\begin{equation}}
\newcommand{\eeq}{\end{equation}}
\newcommand{\bea}{\begin{eqnarray}}
\newcommand{\eea}{\end{eqnarray}}
\newcommand{\bit}{\begin{itemize}}
\newcommand{\eit}{\end{itemize}}
\newcommand{\rf}[1]{(\ref{#1})}
\newcommand\bw{\begin{widetext}}
\newcommand\ew{\end{widetext}}

\newcommand{\KIAA}{\affiliation{Kavli Institute for Astronomy and
Astrophysics, Peking University, Beijing 100871, China}}

\newcommand{\NAOC}{\affiliation{National Astronomical Observatories,
Chinese Academy of Sciences, Beijing 100012, China}}

\begin{document}

\title{Static spherical vacuum solutions in the bumblebee gravity model}

\author{Rui Xu}\email[Corresponding author: ]{xuru@pku.edu.cn}\KIAA
\author{Dicong Liang}\KIAA
\author{Lijing Shao}\email[Corresponding author: ]{lshao@pku.edu.cn}\KIAA\NAOC

\date{\today}

\begin{abstract} 
The bumblebee gravity model is a vector-tensor theory of gravitation where the
vector field nonminimally couples to the Ricci tensor. By investigating the
vacuum field equations with spherical symmetry, we find two families of
black-hole (BH) solutions in this model: one has a vanishing radial component of
the vector field and the other has a vanishing radial component of the Ricci
tensor.  When the coupling between the vector field and the Ricci tensor is set
to zero, the first family becomes the Reissner-Nordstr\"om solution while the
second family degenerates to the Schwarzschild solution with the vector field
being zero. General numerical solutions in both families are obtained for
nonzero coupling between the vector field and the Ricci tensor. Besides BH
solutions, we also reveal the existence of solutions that have a nonvanishing
$tt$-component of the metric on the supposed event horizon where the
$rr$-component of the metric diverges while the curvature scalars are finite.
These solutions are not supported by existing observations but present certain
properties that are of academic interests.  We conclude the study by putting the
BH solutions into tests against the Solar-System observations and the images of
supermassive BHs.

\end{abstract}

\maketitle
\allowdisplaybreaks  

\section{Introduction}
\label{sec:intro}

Einstein's theory of general relativity (GR) and Maxwell's theory of
electromagnetism (EM) are the two representative classical field theories in
physics. A straightforward combination of them, i.e., the Einstein-Maxwell
theory, provides the simplest unification of the gravitational interaction and
the EM interaction at the classical level, and serves as the prototype of
vector-tensor theories in which vector fields are employed to complement the
gravitational interaction described conventionally by the metric tensor. The
action of the Einstein-Maxwell theory reads
\bea
S = \int d^4x \sqrt{-g} \left( \frac{1}{2\ka} R - \frac{1}{4} F^\mn F_\mn \right) + S_{\rm m} ,
\label{actionEM}
\eea   
where $g$ is the determinant of the metric $g_{\mu\nu}$, the constant $\ka$ is
$8\pi G$ with $G$ being the gravitational constant, $R$ is the Ricci scalar,
$F_\mn := D_\mu A_\nu - D_\nu A_\mu$ is the field strength of the vector field
$A_\mu$  with $D_\mu$ being the covariant derivative, and $S_{\rm m}$ represents
the action of matter. Without identifying the vector field as the EM vector
potential, the action in Eq.~\rf{actionEM} was first used by Will and Nordtvedt
to illustrate possible preferred-frame effects in gravity \cite{Will:1972zz}.
The point is that if the vector field $A_\mu$ possesses a nonzero background
configuration, then the spacetime is anisotropic and hence has preferred frames.

Preferred frames violate Lorentz symmetry. The motivation of such a
violation from string theory and the implications of it in low-energy effective
field theories were first studied by Kosteleck\'y and Samuel 
\cite{Kostelecky:1988zi, Kostelecky:1989jp}. A generic framework, called the
Standard-Model Extension (SME), has been developed to systematically incorporate
all possible Lorentz-violating couplings into the actions of the Standard Model
of particle physics and GR \cite{Colladay:1996iz, Colladay:1998fq,
Kostelecky:2003fs, Kostelecky:2009zp, Kostelecky:2011gq, Kostelecky:2013rta,
Kostelecky:2018yfa}, and to calculate a number of predictions that can be tested
in modern high-precision experiments \cite{Kostelecky:2008ts, Tasson:2016xib}.
The primary Lorentz-violating coupling in the gravitational sector of the SME
takes the form $s^\mn R_\mn$, where $s^\mn$ is a tensor field possessing a
nonzero background configuration and $R_\mn$ is the Ricci tensor. Because the
background configuration of $s^\mn$ defines preferred frames in
general, this coupling violates Lorentz symmetry. The SME focuses on general
properties and consequences caused by the coupling term, but asks little about
the dynamics of the field $s^\mn$.  To fill the gap between the SME and specific
gravitational theories having the Lorentz-violating coupling, the action in
Eq.~\rf{actionEM} was generalized to \cite{Kostelecky:2003fs, Bailey:2006fd}
\bea
S &=& \int d^4x \sqrt{-g} \left( \frac{1}{2\ka} R + \frac{\xi}{2\ka} B^\mu B^\nu R_\mn - \frac{1}{4} B^\mn B_\mn - V \right) 
\nonumber \\
& & + S_{\rm m} ,
\label{actionB}
\eea      
where $B^\mu$ is a dynamical vector field sometimes called the bumblebee field,
and the generalized vector-tensor theory (\ref{actionB}) is called the bumblebee
gravity model \cite{Kostelecky:2003fs}.

Apart from replacing the vector field $A^\mu$ by $B^\mu$ (correspondingly
$F_\mn$ by $B_\mn:= D_\mu B_\nu - D_\nu B_\mu$), the important changes are
adding (i) the coupling term $B^\mu B^\nu R_\mn$ controlled by the constant
$\xi$ to resemble the SME term $s^\mn R_\mn$, and (ii) the potential $V$ that
takes its extremum when the bumblebee vector field $B^\mu$ acquires a certain
background configuration. By considering the linearized, perturbative bumblebee
model where the deviation of the metric from the Minkowski metric
$\eta_{\mu\nu}$ and the deviation of the bumblebee vector field from its
background configuration are small, conclusions made in the general framework of
the SME are verified by and contrasted with results in the bumblebee model
\cite{Kostelecky:2003fs, Bluhm:2004ep, Bluhm:2005uj, Bailey:2006fd,
Bluhm:2007bd, Bluhm:2008yt, Liang:2022hxd}.

Simply speaking, the bumblebee model is an essential example theory for studying
Lorentz violation in gravity as its specified action provides information on the
dynamics of the auxiliary field that breaks Lorentz symmetry, which is otherwise
not contained in the general framework of the SME. But it is more than that. In
fact, the action in Eq.~\rf{actionB} without the potential $V$ has been studied
by \citet{Hellings:1973zz} as an alternative to GR in the context of the
parametrized post-Newtonian (PPN) formalism as well as the cosmological
solutions. At the end of their work, they brought up the proposal of identifying
the vector field as the EM vector potential.  A similar possibility was
suggested by \citet{Bluhm:2004ep} in the framework of the SME, where EM waves
are the Nambu-Goldstone modes of the bumblebee vector field when Lorentz
symmetry is spontaneously broken due to a background bumblebee field resulting from the potential $V$. The idea of replacing the Einstein-Maxwell theory
with the bumblebee model for unifying gravity and the EM theory is attractive.

Compared with the Einstein-Maxwell theory, the bumblebee model is expected to
produce new and maybe even eccentric phenomena when the vector field interacts
with gravity through the coupling term $B^\mu B^\nu R_\mn$. The gravitational
field is optimally to be strong for such phenomena to be detectable.  Therefore,
strong-field solutions without applying the perturbation approach, as was done
in the SME or the PPN formalism, should be considered. Black holes (BHs) are
ideal strong-field systems to study, for not only the absence of matter
simplifies the field equations, but also they have been detected with both
gravitational waves (GWs) \cite{LIGOScientific:2018mvr, LIGOScientific:2020ibl,
LIGOScientific:2021djp} and EM waves \cite{EventHorizonTelescope:2019dse,
EventHorizonTelescope:2022xnr} so that unprecedented tests can be performed
using the rapidly developing technology of multimessenger
astronomy~\cite{EventHorizonTelescope:2022xqj}.

In the bumblebee model, an analytical solution that is very close to the
Schwarzschild spacetime with the bumblebee field having only a nonvanishing
radial component has been reported by \citet{Casana:2017jkc}. In our work, we
substantially extend the study and find spherical solutions with a nonvanishing
temporal component of the bumblebee field. In fact, as suggested by the EM-like
kinetic term in the action, the radial component of the bumblebee field is
nondynamic, and can be eliminated from the final set of equations.  It turns out
there are two families of solutions: one solved from a group of two second-order
ordinary differential equations (ODEs), and one solved from a group of three
second-order ODEs. The presence of the temporal component of the bumblebee field
is vital for both families of solutions as it alters the behavior of the metric
near the BH event horizon radically. Most amazingly, it makes solutions with a
nonvanishing $g_{tt}$ at the event horizon possible.   

We find that it is necessary to distinguish the BH solutions where $g_{tt}$
vanishes at the event horizon from the solutions having a nonvanishing $g_{tt}$
at the event horizon, for geodesics in the spacetime of the latter solutions
remarkably bounce back on the event horizon. For this reason, those solutions
with a nonvanishing $g_{tt}$ at the event horizon are called {\it{compact
hills}} (CHs) in our work. The bouncing-back behavior of geodesics has no
experimental or observational evidence in gravity phenomena yet. Nevertheless,
academically interesting features of the CH solutions are discussed, including
the GW echoes~\cite{Cardoso:2017cqb}.

For the BH solutions, we put them into tests against Solar-System observations
and the shadow images of supermassive BHs achieved by the Event Horizon
Telescope (EHT) Collaboration. As the two families of solutions have different
numbers of free parameters, constraints on their parameter spaces are different.
In one family, the BHs are characterized by two parameters, namely the mass and
the bumblebee charge, so bounds on the bumblebee charge of the Sun and of the
supermassive BHs are obtained. In the other family, the BHs have five free
parameters where three of them are in the metric functions and two of them are
in the bumblebee field. The Solar-System observations happen to only depend on
the metric parameters for the solutions in this family, so bounds on the metric
parameters are obtained while leaving the two bumblebee parameters
unconstrained. The parameter space excluded in our work is limited, especially
for the second family of solutions. Future study on the GWs in the spacetime of
the BH solutions might yield tighter or complementary constraints.

The organization of this paper is as follows. We start with setting up the
equations in Sec.~\ref{sec:IIa}. Then an analytical solution  is presented in
Sec.~\ref{sec:IIb}. General numerical solutions are discussed in
Sec.~\ref{sec:IIc}, and a detailed discussion about the CH solutions is made in
Sec.~\ref{consequences}. In Sec.~\ref{sec:IIIa}, the BH solutions are tested by
considering Solar-System observations, while in Sec.~\ref{sec:IIIb}, the test
is done by considering the results from the shadow images of  supermassive BHs.
Finally, we summarize our findings and give a brief outlook for the directions
worthy of further study in Sec.~\ref{sec:sum}. The appendices list equations to
supplement  the  main text.

In the remaining of the paper, equations are written in the geometrized unit
system where $G = c = 1$. The sign convention of the metric is $(-, +, +, +)$.

\section{Static spherical solutions in vacuum}
\label{sec:II}

\subsection{Field equations}
\label{sec:IIa}

The field equations are obtained by taking variations with respect to the metric
and the bumblebee field in Eq.~\rf{actionB}. Under the assumption that the
bumblebee field does not couple to conventional matter, the field equations can
be written as
\bea
&& G_\mn = \ka \left( T_{\rm m} \right)_\mn +  \ka \left( T_{B}\right)_\mn , 
\nonumber \\
&& D^\mu B_{\mn} - 2B_\nu \frac{dV}{d(B^\la B_\la)} + \frac{\xi}{\ka} B^\mu R_\mn  = 0,
\eea
where $\left( T_{\rm m} \right)_\mn$ is the energy-momentum tensor for
conventional matter, and the contribution of the bumblebee field to the Einstein
equations is
\bea
\left( T_{B}\right)_\mn &=& \frac{\xi}{2\ka} \Big[ g_\mn B^\al B^\be R_{\al\be} - 2 B_\mu B_\la R_\nu^{\pt\nu \la} - 2 B_\nu B_\la R_\mu^{\pt\mu \la} - \Box_g ( B_\mu B_\nu ) 
\nonumber \\
&& - g_{\mn} D_\al D_\be ( B^\al B^\be ) + D_\ka D_\mu \left( B^\ka B_\nu \right) + D_\ka D_\nu ( B_\mu B^\ka )   \Big]
\nonumber \\
&& + B_{\mu\la} B_\nu^{\pt\nu \la} - g_\mn \left( \frac{1}{4} B^{\al\be} B_{\al\be} + V \right) + 2 B_\mu B_\nu \frac{dV}{d(B^\la B_\la)} .
\eea 
The d'Alembertian in the curved spacetime is $\Box_g :=g^{\al\be}D_\al D_\be$.
Note that the potential $V$ has been assumed to be a function of the scalar
product $B^\mu B_\mu$.

We are interested in nontrivial background configurations of the bumblebee field
that are compatible with vacuum, namely $\left( T_{\rm m} \right)_\mn = 0$. The
potential $V$ is supposed to take extrema at these background configurations.
Denoting $B_\mu = b_\mu$ as a background bumblebee field, we have 
\bea
\frac{dV}{d(B^\la B_\la)} \Big|_{B_\mu = b_\mu}  = 0 .
\eea   
In addition, the value of $V$ at $B_\mu = b_\mu$ is equivalent to a cosmological
constant. We drop it in the current study of BH solutions. Therefore, the vacuum
field equations are simplified to 
\bea
&& G_\mn =  \ka \left( T_{b}\right)_\mn , 
\nonumber \\
&& D^\mu b_{\mn} + \frac{\xi}{\ka} b^\mu R_\mn  = 0,
\label{fieldeqs2}
\eea
where $b_\mn = D_\mu b_\nu - D_\nu b_\mu$, and 
\bea
\left( T_{b}\right)_\mn &=& \frac{\xi}{2\ka} \Big[ g_\mn b^\al b^\be R_{\al\be} - 2 b_\mu b_\la R_\nu^{\pt\nu \la} - 2 b_\nu b_\la R_\mu^{\pt\mu \la} - \Box_g ( b_\mu b_\nu ) 
\nonumber \\
&& - g_{\mn} D_\al D_\be ( b^\al b^\be ) + D_\ka D_\mu \left( b^\ka b_\nu \right) + D_\ka D_\nu ( b_\mu b^\ka )   \Big]
\nonumber \\
&& + b_{\mu\la} b_\nu^{\pt\nu \la} - \frac{1}{4} g_\mn  b^{\al\be} b_{\al\be}   .
\eea

Focusing on static spherical solutions, we use the metric ansatz
\bea
ds^2 = -e^{2\nu}dt^2 + e^{2\mu} dr^2 + r^2 \left( d\th^2 + \sin^2\th \, d\ph^2 \right) ,
\label{ssmetric}
\eea
and assume the background bumblebee field to be $b_\al = \left( b_t,\, b_r, \,
0, \, 0 \right)$, where the unknowns, $\mu\, , \nu, \, b_t$ and $b_r$, are
functions of $r$. Then the Einstein field equations are nonvanishing for the
diagonal components as well as the $tr$-component, and the vector field equation
has nonvanishing $t$-component and $r$-component. It turns out that the
$tr$-component of the Einstein field equations is guaranteed by the
$r$-component of the vector field equation, leaving 5 equations for the 4
unknowns, $\mu\, , \nu, \, b_t$ and $b_r$. So one of the equations depends on
the others. These equations are displayed in Appendix~\ref{app1}.

\subsection{An analytical solution}
\label{sec:IIb}

Inspired by the solution obtained in Ref.~\cite{Casana:2017jkc}, we checked 
\bea
&& \nu= \frac{1}{2} \ln{\left(1-\frac{2M}{r} \right)} , 
\nonumber \\
&& \mu = \mu_0 - \frac{1}{2} \ln{\left(1-\frac{2M}{r} \right)}, 
\nonumber \\
&& b_t = \la_{0} + \frac{\la_{1}}{r} ,
\label{asol1}
\eea
against Eqs.~\rf{ssefe} and \rf{ssvfe}, or equivalently Eq.~\rf{fieldeqs2} with
the static spherical ansatz, to find that as long as 
\bea
b_r^2 &=& e^{2\mu_{0}} \Bigg[ \frac{1}{\xi} \frac{\left(e^{2\mu_{0}}-1\right)r }{ r-2M} - \frac{\ka \la_1^2}{3\xi M (r-2M)}
\nonumber \\
&& + \frac{\la_1^2 (2r-M) + 6\la_0\la_1 Mr + 6 \la_0^2 M^2 r}{3M (r-2M)^2} \Bigg] ,
\label{asol2}
\eea 
Eq.~\rf{asol1} is a solution to the field equations. It is an analytical solution
characterized by 4 integral constants, $\mu_0, \, M, \, \la_{0}$ and $\la_{1}$.
The 2-parameter solution obtained in Ref.~\cite{Casana:2017jkc} simply
corresponds to $\la_{0}=\la_{1}=0$. We also notice that the 3-parameter solution corresponding to $\mu_0=0$ has been previously obtained in Ref.~\cite{Fan:2017bka}.

The following two observations can be made based upon the solution given by
Eqs.~\rf{asol1} and \rf{asol2}.
\begin{enumerate} 
\item The Schwarzschild metric corresponding to $\mu_0 = 0$ is no longer
necessarily accompanied by a vanishing bumblebee field. Among the valid choices
of $\la_0$ and $\la_1$ that give nonnegative $b_r^2$, an intriguing choice is
$\la_1 = -2M \la_0$, which simplifies Eq.~\rf{asol2} to 
\bea
b_r^2 &=& \frac{2}{3} \left(1-\frac{2\ka}{\xi} \right) \frac{ \la_0^2 M }{r-2M } ,
\eea 
for $\mu_0 =0$. As we require $b_r^2 \ge 0$ outside the event horizon, this
choice of $\la_1$ can only be made for $\xi \ge 2\ka$. For the special case of
$\xi=2\ka$, the Schwarzschild metric is a solution to the field equations
together with a nontrivial bumblebee field that has only a temporal component
$b_t \propto 1 - 2M/r$. 
 
\item The Minkowski metric is recovered with $\mu_0=0$ and $M=0$. The limit of
Eq.~\rf{asol2} exists if $\la_1 \propto \sqrt{M}$ while $M \rightarrow 0$. That
is to say, the Minkowski metric can be accompanied by a nontrivial bumblebee
field that has a radial component 
\bea
b_r^2 = \frac{1}{3} \left( 2 - \frac{\ka}{\xi} \right) \frac{C}{r}, 
\eea
where $C$ being the ratio $\la_1^2/M$ has the dimension of length, and might
represent the length scale of possible Lorentz violation in an underlying
theory.
\end{enumerate}

\subsection{General numerical solutions}
\label{sec:IIc}

General solutions to Eqs.~\rf{ssefe} and \rf{ssvfe} are obtained numerically in
two cases enumerated in Appendix~\ref{app1} according to the two factors in the
$r$-component of the vector equation: (i) $b_r = 0$; (ii) $R_{rr} = 0$. For the
first case, the system of equations reduce to two coupled second-order ODEs,
indicating a family of solutions determined by 4 parameters. For the second
case, the system of equations reduce to three coupled second-order ODEs,
indicating a family of solutions determined by 6 parameters.  Both families of
solutions turn out to have the asymptotic expansions 
\bea
\nu = \sum\limits_{n=1}^{n=\infty} \frac{\nu_{n}}{r^n}, \nonumber \\
\mu = \sum\limits_{n=0}^{n=\infty} \frac{\mu_{n}}{r^n}, \nonumber \\
b_t = \sum\limits_{n=0}^{n=\infty} \frac{\la_{n}}{r^n}.   
\label{asyexp}
\eea
After substituting them into the ODEs to find the recurrence relations for the
expansion coefficients $\nu_n, \, \mu_n$ and $\la_n$, these expansions can be
utilized to assign the initial values to start the numerical integrations from
large enough $r$.  Note that a constant term in the expansion of $\nu$ is
unnecessary as it amounts to a change of the time coordinate. Therefore, the
number of parameters characterizing the solutions is actually three for the
first family ($b_r=0$) and 5 for the second family ($R_{rr}=0$).  

The recurrence relations for the first few coefficients are given in
Appendix~\ref{app2}.  The conclusion is that for the case of $b_r = 0$, the
solutions must have $\mu_0=0$ and are fixed once the three free coefficients
$\mu_1, \, \la_0$ and $\la_1$ are specified, and that for the case of
$R_{rr}=0$, there are 5 free coefficients, $\mu_0, \, \mu_1, \, \mu_2, \, \la_0$
and $\la_1$, describing the solutions. We point out that the analytical solution
given by Eqs.~\rf{asol1} and \rf{asol2} belongs to the second case and is
recovered when taking $\mu_2=\mu_1^2=M^2$.

Before further discussing the numerical solutions, let us relate the parameters
$\mu_1$ and $\la_1$ to the mass and the bumblebee charge of the spacetime. With
$\mu_0 = 0$, the Arnowitt-Deser-Misner (ADM) mass of the spacetime can be
calculated and it is $\mu_1$ \cite{Arnowitt:1962hi}. However, the definition of
the ADM mass does not apply to the solutions with $\mu_0 \ne 0$, when the metric
is not asymptotically flat. Therefore we adopt the Komar mass as a measure of
the spacetime mass when $\mu_0\ne 0$. The Komar mass is defined by the conserved
current associated with the time-translation Killing vector $K^\mu = (1, 0, 0,
0)$. The conserved current is \cite{Poisson:2009pwt} 
\bea
J_M^\mu : = K_\nu R^\mn =  D_\nu D^\mu K^\nu ,
\eea    
where the curvature identity $[D_\mu, D_\nu] K^\al = R^\al_{\pt\al\la\mn} K^\la$ and the Killing equation $D_\mu K_\nu + D_\nu K_\mu = 0$ have been used. The Komar mass then can be calculated as
\bea
M_{\rm K} &:=& -\frac{1}{4\pi} \int d^3x \sqrt{-g} \, J_M^t 
\nonumber \\
&=& -\frac{1}{4\pi} \lim\limits_{r\rightarrow \infty} \iint d\th d\ph \sqrt{-g} \, D^t K^r 
\nonumber \\
&=& e^{-\mu_0} \mu_1.
\eea
It is evident that the Komar mass reduces to the ADM mass when $\mu_0 = 0$.  The
bumblebee charge can be defined in a similar way. Instead of using a Killing
vector, the current 
\bea
J_Q^\mu := \frac{\xi}{\ka} b_\nu R^\mn = - D_\nu b^{\nu\mu}
\eea  
is automatically conserved as $b_\mn$ is antisymmetric. Therefore the bumblebee
charge having the dimension of length can be defined as
\bea
Q &: =& -\frac{ 1 }{4\pi} \sqrt{\frac{\ka}{2}} \int d^3x \sqrt{-g} \, J_Q^t 
\nonumber \\
&=& -\frac{1}{4\pi} \sqrt{\frac{\ka}{2}} \lim\limits_{r\rightarrow \infty} \iint d\th d\ph \sqrt{-g} \, b^{tr} 
\nonumber \\
&=& \sqrt{\frac{\ka}{2}} \, e^{-\mu_0} \la_1 , 
\eea
where the factor $\sqrt{\ka/2}$ has been chosen so that the Reissner-Nordstr\"om
metric in Eq.~\rf{rnmetric} can be recovered when $\xi=0$.

Back to the numerical results, as we integrate the system of ODEs from a large
radius to $r=0$, there exist three types of solutions. The one that we are
interested in has $g_{rr}=e^{2\mu}$ diverging at a certain radius $r=r_h$. The
other two types that we will ignore are solutions of naked singularity and
solutions with $g_{tt}=-e^{2\nu}$ diverging at a finite radius. Concentrating on
the interesting solutions that have $g_{rr}$ diverging at a finite radius
$r=r_h$, the BH solutions are selected with two extra conditions: (i) vanishing
$g_{tt}$ at $r_h$ and (ii) finite curvature scalars at $r_h$. It turns out that
the finiteness of the curvature scalars at $r_h$ is trivially satisfied by all
the solutions with $g_{rr}$ diverging at $r_h$. But among these solutions, it
turns out that not all of them have $g_{tt}=0$ at $r_h$. In fact, for the
solutions in the first family ($b_r=0$), most of them have nonvanishing $g_{tt}$
at $r_h$.    

To comprehend what happens near $r_h$, we seek expansions of the variables in
terms of $\de: = r-r_h$. It is after a careful investigation of the numerical
solutions that we find the variables taking the following expansions near $r_h$,
\bea
&& g_{tt} = -e^{2\nu} = -\sum\limits_{n=0}^{\infty} N_{1n} \, \de^n -\sqrt{\de} \sum\limits_{n=0}^{\infty} N_{2n} \, \de^n,
\nonumber \\
&& g_{rr} = e^{2\mu} = \frac{1}{\de}\sum\limits_{n=0}^{\infty} M_{1n} \, \de^n + \frac{1}{\sqrt{\de}} \sum\limits_{n=0}^{\infty} M_{2n} \, \de^n,
\nonumber \\
&& b_t = \sum\limits_{n=0}^{\infty} L_{1n} \, \de^n + \sqrt{\de} \sum\limits_{n=0}^{\infty} L_{2n} \, \de^n,
\label{hexp}
\eea
where $\big\{ N_{1n}, \, N_{2n},  \, M_{1n}, \, M_{2n}, L_{1n}, \, L_{2n}
\big\}$ is the set of expansion coefficients. The recurrence relations can be
obtained by substituting the expansions into the ODEs and setting each order of
$\de$ to zero. The recurrence relations for the first few coefficients are given
in Appendix~\ref{app3}. The most important fact we learn is that the coefficient
$N_{10}$ is closely related to the coefficients for the terms involving
$\sqrt{\de}$ in the following way. If $N_{10}=0$, then $N_{2n}=M_{2n}=L_{2n}=0$,
and vice versa.

The solutions with $g_{rr}$ diverging at a finite radius $r_h$, namely those can
be expanded in the form of Eq.~\rf{hexp}, thus can be divided into two
categories according to whether $N_{10}$ is zero. The BH solutions have
$N_{10}=0$, while the solutions with $N_{10}\ne 0$, for their unexpected
property of bouncing geodesics at $r_h$, which will be discussed in
Sec.~\ref{consequences}, will be dubbed CHs in the remaining of the paper. Note
that both the BH solutions and the CH solutions have finite curvature scalars at
$r_h$ (see Appendix~\ref{app4}). 

\begin{figure*}
 \includegraphics[width=\linewidth]{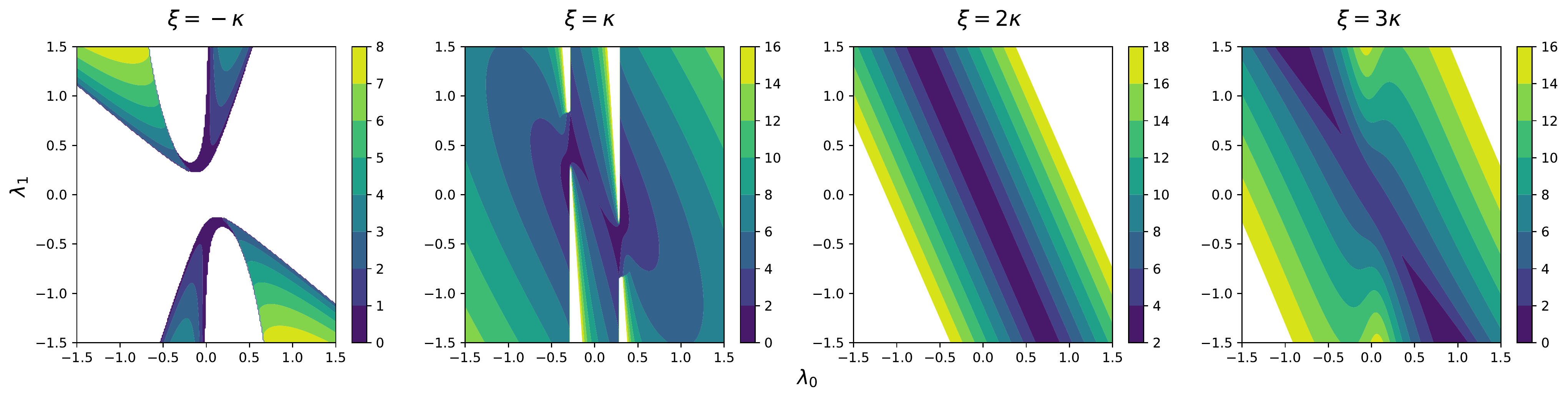}
 \caption{Contour plots for $r_h$ on the $\la_0$-$\la_1$ plane for the
 solutions in the first family ($b_r=0$). The asymptotic parameter $\mu_1$ is
 used as the length unit in the geometrized unit system. The blank region in the
 first plot consists of uninteresting solutions, namely naked singularities and
 solutions with singular $g_{tt}$.  Notice that when $\xi=2\ka$, metric
 solutions are identical given $\la_0$ and $\la_1$ on a line $\la_1 = -2\la_0 +
 {\rm constant}$. The line $\la_1 = -2\la_0$ gives the Schwarzschild metric
 specially.
 }
\label{fig1}
\end{figure*}
\begin{figure*}
 \includegraphics[width=\linewidth]{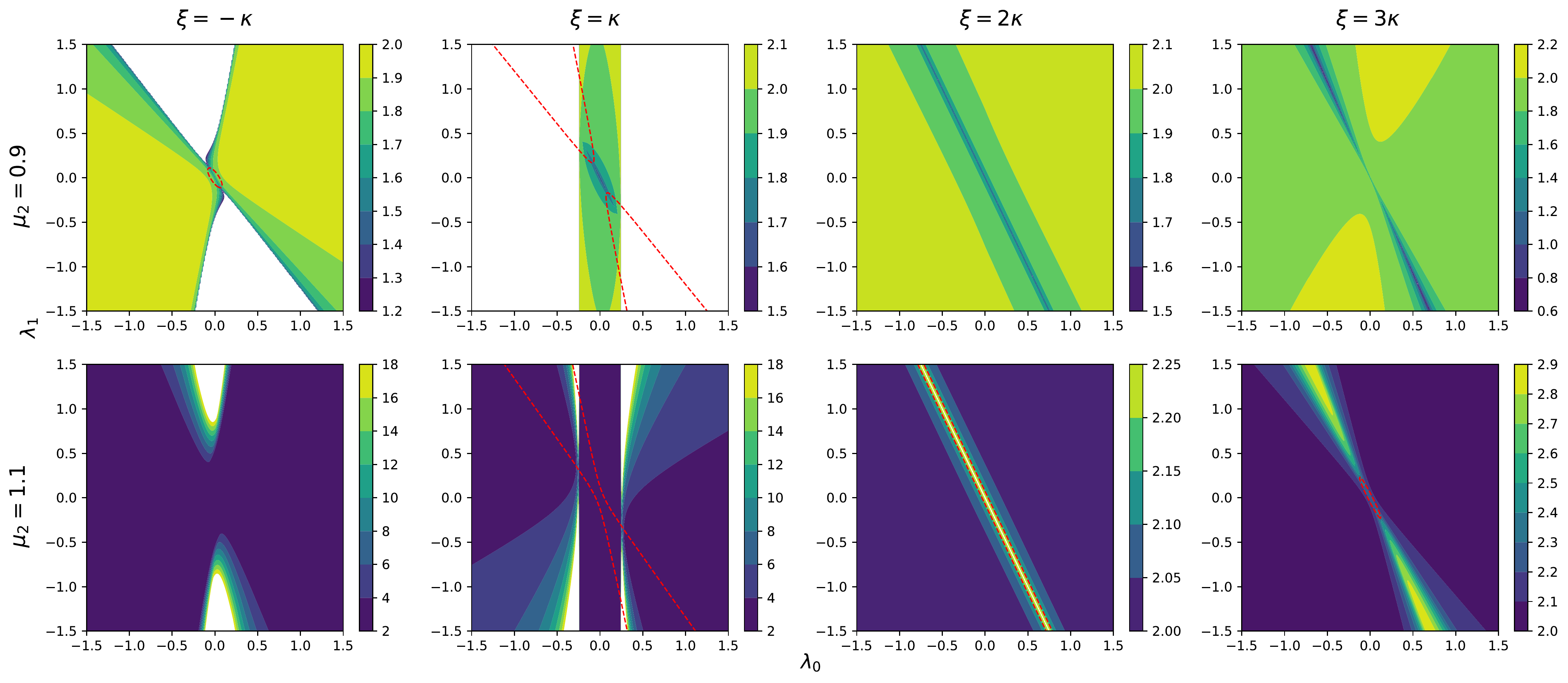}
 \caption{Contour plots for $r_h$ on the $\la_0$-$\la_1$ plane for the
 solutions in the second family ($R_{rr}=0$). We have taken $\mu_0=0$ and used
 $\mu_1$ as the length unit in the geometrized unit system. The blank regions in
 the two plots in the first row consist of naked singularities. The red lines
 are boundaries of the inequality \rf{brsqcon}, which is a requirement of
 $b_r^2\ge 0$ at infinity. Among the 5 plots with the red lines, the region for
 the inequality to hold contains the origin $\la_0=\la_1=0$ only for the second
 plot in the first row.}
\label{fig2}
\end{figure*}

Figures \ref{fig1} and \ref{fig2} show how $r_h$ depends on the asymptotic
parameters $\la_0$ and $\la_1$. We take the asymptotic parameter $\mu_1$ as the
length unit in the numerical calculations. For the first family of solutions
($b_r=0$), a solution is obtained once the values of the coupling constant $\xi$
and the asymptotic parameters $\la_0$ and $\la_1$ are specified properly.
Sufficient amount of solutions obtained by varying $\la_0$ and $\la_1$ then
generate the contour plots of $r_h$ on the $\la_0$-$\la_1$ plane as shown in
Fig.~\ref{fig1}. For the second family of solutions ($R_{rr}=0$), with $\mu_1$
being the unit, a solution is obtained when the asymptotic parameters $\mu_0,\,
\mu_2, \, \la_0, \, \la_1$, as well as the coupling constant $\xi$ are
specified properly.  To generate Fig.~\ref{fig2}, we have taken $\mu_0=0$ and
two representative values for $\mu_2$. Solutions with nonzero but small $\mu_0$
turn out to produce similar shapes of the contour plots, while solutions with
$\mu_0$ deviating from zero significantly are uninteresting to us as $\mu_0$ has
been constrained to be smaller than $10^{-12}$ using the advance of perihelia of
the Solar-System planets in Ref.~\cite{Casana:2017jkc}.

\begin{figure*}
 \includegraphics[width=\linewidth]{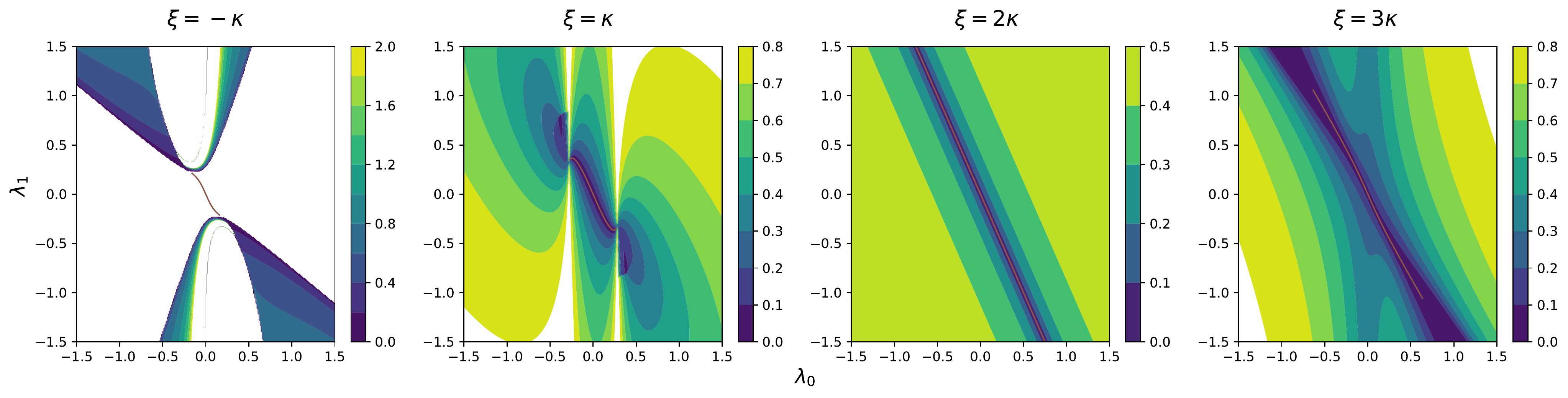}
 \caption{Contour plots for $N_{10}=-g_{tt}(r_h)$ on the $\la_0$-$\la_1$ plane
 for the solutions in the first family ($b_{r}=0$). The BH solutions,
 corresponding to $N_{10}=0$, are located along the brown lines. The colored
 contours with $N_{10}>0$ are solutions dubbed ``CHs'' in our work.}
\label{fig3}
\end{figure*}

\begin{figure*}
 \includegraphics[width=\linewidth]{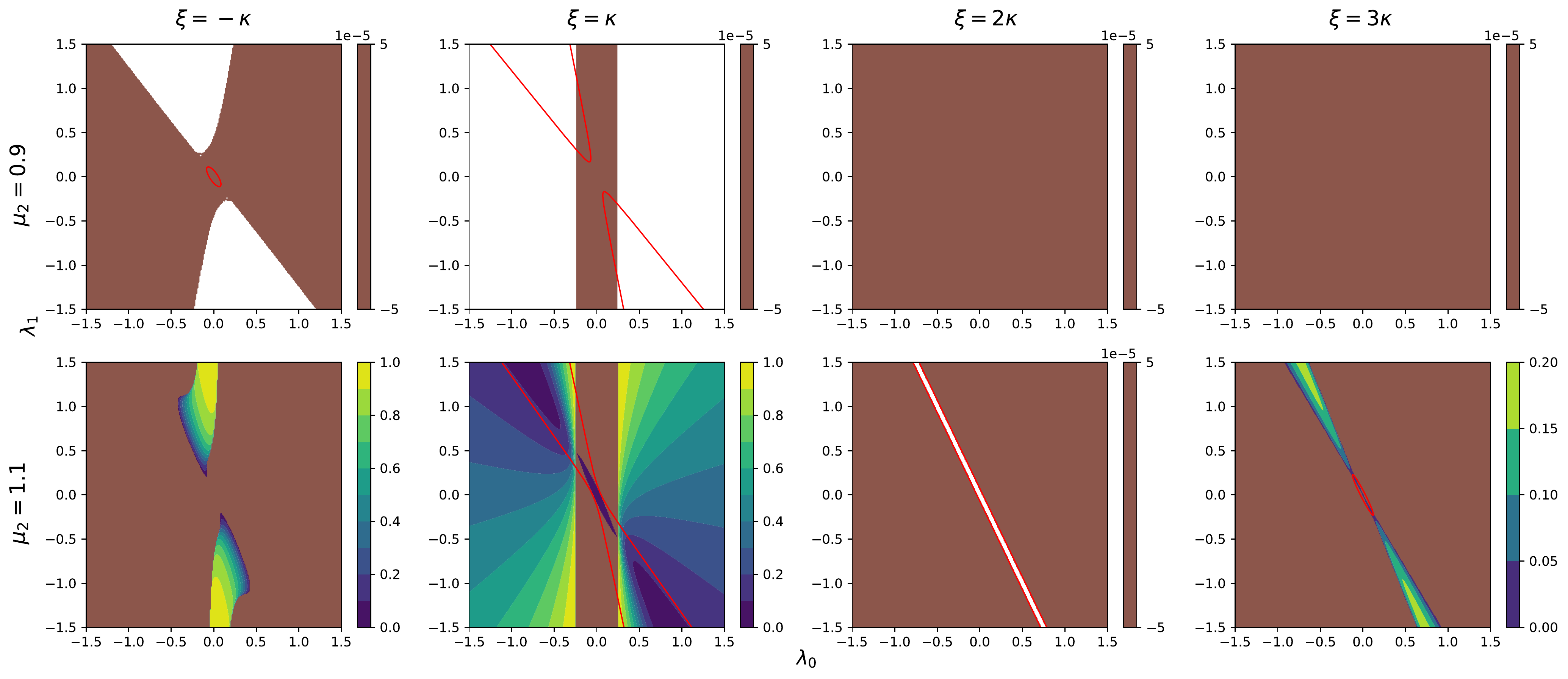}
 \caption{Contour plots for $N_{10}=-g_{tt}(r_h)$ on the $\la_0$-$\la_1$ plane
 for the solutions in the second family ($R_{rr}=0$). The BH solutions,
 corresponding to $N_{10}=0$, form the regions colored in brown. We have checked
 that our numerical integrations have errors less than $5 \times 10^{-5}$. The
 contours in other colors represent solutions with $N_{10}>0$. These solutions,
 dubbed ``CHs'', do not exist for $\xi=2\ka$ or when $\mu_2 \le \mu_1^2$. The
 red lines have the same meaning as in Fig.~\ref{fig2}.}
\label{fig4}
\end{figure*}

Figures \ref{fig3} and \ref{fig4} show how $N_{10}$ depends on $\la_0$ and
$\la_1$, using the same solutions producing Figs.~\ref{fig1} and \ref{fig2}. The
values of the asymptotic parameters $\la_0$ and $\la_1$ for BH solutions and for
CH solutions are conveniently revealed in Figs.~\ref{fig3} and \ref{fig4}. For
the first family of solutions, the requirement of $N_{10}=0$ turns out to force
$L_{10}$ to vanish, which acts as an extra constraint on the asymptotic
parameters $\la_0$ and $\la_1$. Therefore, the BH solutions in the first family
only exist along a single line on the $\la_0$-$\la_1$ plane for a given $\xi$,
leaving the majority of the valid $\la_0$ and $\la_1$ to generate the CH
solutions characterized by $N_{10}\ne 0$. For the second family of solutions,
the requirement of $N_{10}=0$ forces no extra constraint on $\la_0$ and $\la_1$,
so the BH solutions exist on the two-dimensional plane of $\la_0$ and $\la_1$.
In fact, as shown in Fig.~\ref{fig4}, there is no CH solution for any values of
$\la_0$ and $\la_1$ as long as one takes $\mu_2 \le \mu_1^2$.   


\begin{figure}
 \includegraphics[width=\linewidth]{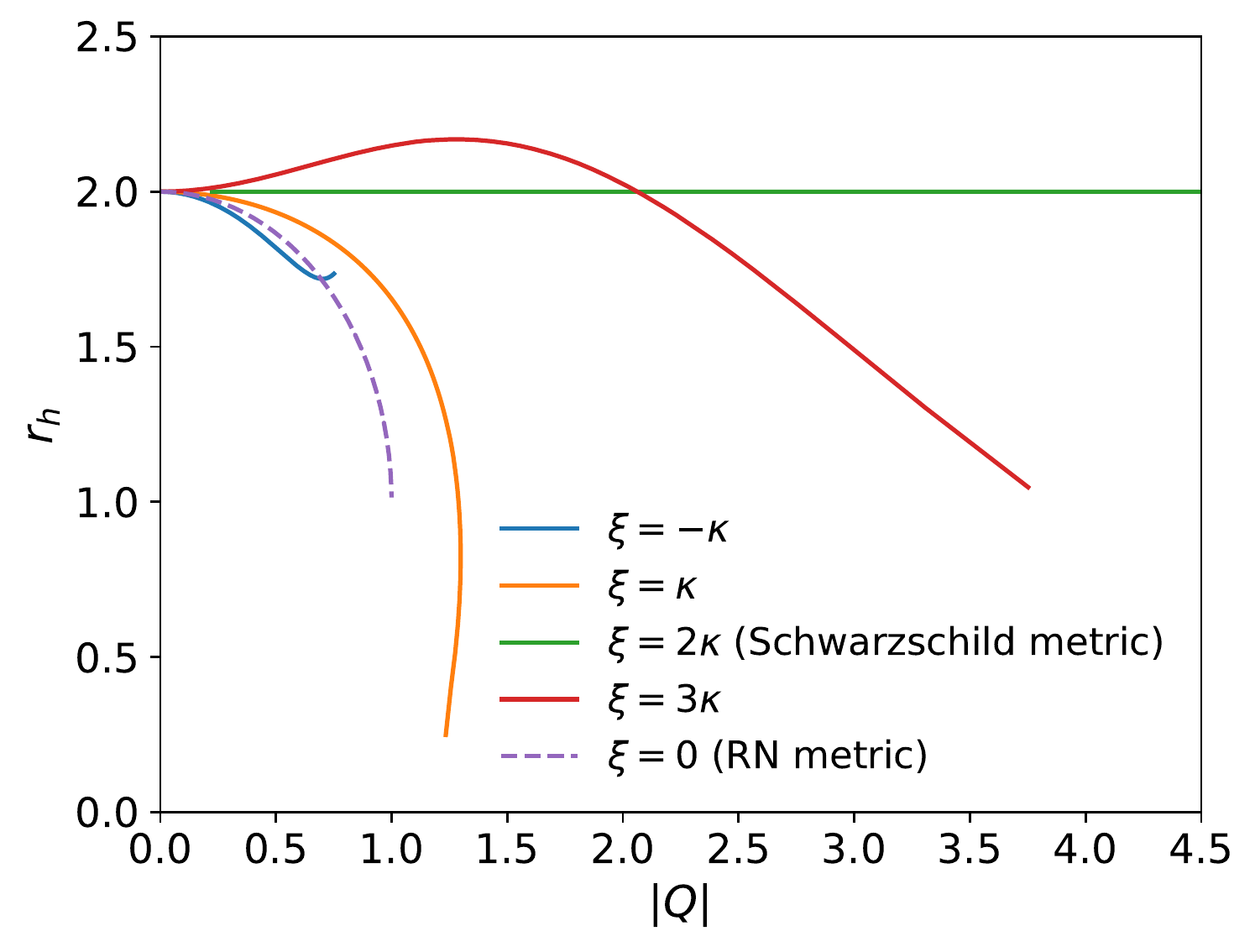}
 \caption{Event horizon $r_h$ versus bumblebee charge $Q$ for the BH solutions
 in the first family ($b_r=0$). The unit for $r_h$ and $Q$ is $\mu_1$, which is
 the mass of the BH. }
\label{fig5}
\end{figure}

Comparing the BH solutions in the first family ($b_r=0$) with the charged BH
solution in GR, i.e., the Reissner-Nordstr\"om metric, is worthwhile. The
asymptotic expansions of the metric functions $\mu$ and $\nu$ given by the
recurrence relations in Eq.~\rf{case1asyrec} include the Reissner-Nordstr\"om
metric as a special case for $\xi=0$, when the bumblebee model recovers the
Einstein-Maxwell theory. This suggests that the bumblebee BH solutions in the
first family, characterized by the two parameters, $M=\mu_1$ and $Q=\sqrt{\ka/2}
\, \la_1$, can be viewed as a generalization of the Reissner-Nordstr\"om metric.
A plot of $r_h$ versus $Q$ in unit of $\mu_1$ as shown in Fig.~\ref{fig5}
demonstrates how the family of solutions change with the coupling constant
$\xi$. We would like to direct the readers' attention to two special values of
$\xi$, namely $\xi=0$ and $\xi=2\ka$. When $\xi<0$, the maximal charge is less
than $M$, and when $\xi>0$, the maximal charge is greater than $M$. When
$\xi<2\ka$, the maximal radius of the event horizon is $2M$ (Schwarzschild metric), and
when $\xi>2\ka$, the maximal radius of the event horizon is greater than $2M$. We notice that when $\xi>2\ka$ or $\xi<0$ there exist two black-hole solutions with the same $r_h$ but different $Q$. The two black holes are certainly observationally distinguishable as the difference in $Q$ makes their metric functions different. In fact, when we calculate the shadow of the black holes in Sec.~\ref{sec:IIIb}, we see that they have different shadow radii (see Fig.~\ref{fig10}).  
A perhaps more interesting but puzzling observation is that when $\xi=\ka$, $r_h$ has two values given $Q$ near its maximum. When this happens, we suspect that the black hole with the smaller $r_h$ might be unstable. Further study on quasi-normal modes of the black holes found here will shed light on this point.    

\begin{figure*}
 \includegraphics[width=0.8\linewidth]{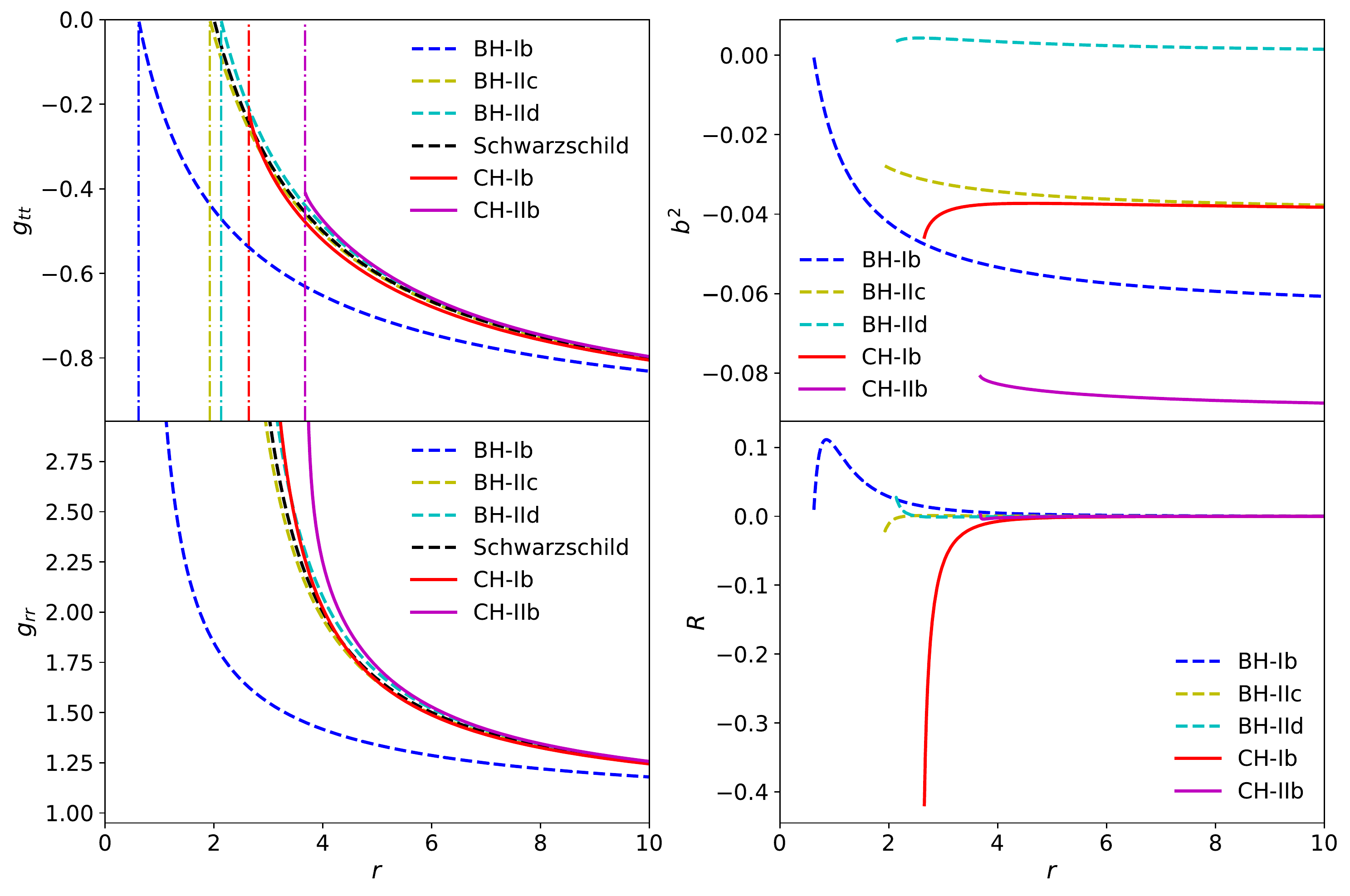}
 \caption{Profiles of several representative solutions. Vertical lines are
 drawn in the plot of $g_{tt}$ to indicate the event horizon of each solution.
 While the Schwarzschild solution with a zero bumblebee field has vanishing
 $b^2:=g^\mn b_\mu b_\nu$ and the Ricci scalar $R$, the two scalars are nonzero
 for general solutions in the bumblebee model as shown in the right panels.   }
\label{fig6}
\end{figure*}

We end the general discussion on the solutions by tabulating several
representative numerical solutions for further reference. Table \ref{solcase1}
lists 4 BH solutions and 4 CH solutions in the first family. Table
\ref{solcase2} lists 8 BH solutions and 3 CH solutions in the second family. To
have a brief picture about how different from the Schwarzschild metric these
solutions are, the metric functions, as well as the scalar $b^2:=g^\mn b_\mu
b_\nu$ and the Ricci scalar $R$, are plotted in Fig.~\ref{fig6}, for several of
the representative solutions.

\begin{table*}
\caption{Representative solutions in the first family ($b_r=0$). The asymptotic
parameter $\mu_1$ is used as the length unit in the geometrized unit system. In
the columns of ``Event horizon behavior'', the leading terms of the expansions
in Eq.~\rf{hexp} are given. }
\renewcommand{\arraystretch}{2}
\begin{tabular}{m{1.3cm}m{1.0cm}m{1.4cm}m{2.2cm}m{1.4cm}m{3.1cm}m{2.2cm}m{3cm}}
\hline\hline
 Solution & $\xi$ & \multicolumn{2}{l}{Asymptotic parameters} & \multicolumn{4}{c}{Event horizon behavior} \\
\hline
 &  & $\la_0$ & $\la_1$ & $r_h$ & $-g_{tt}$ & $g_{rr}$ & $b_t$  \\
\hline
BH-Ia & $-\ka$  & $0.1389$ & $-0.1974$  & $1.719$ & $0.4913 \, \de$ & $\frac{3.500}{\de}$ & $0.1254\,\de$  \\
CH-Ia & $-\ka$  & $0.2$ & $-0.25$  &  $1.079$ & $0.04675-0.05157\sqrt{\de}$ & $\frac{0.6751}{\de}-\frac{0.05777}{\sqrt{\de}}$ & $-0.07031+0.07584\sqrt{\de}$  \\
\hline
BH-Ib & $\ka$   & $0.2568$ & $-0.3637$  & $0.6168$ & $0.6754 \, \de$ & $\frac{0.8299}{\de}$ & $0.2454\, \de$  \\
CH-Ib & $\ka$   & $0.2$ & $-0.25$  &  $2.624$ & $0.1839+0.2205\sqrt{\de}$ & $\frac{0.4535}{\de}+\frac{1.080}{\sqrt{\de}}$ & $0.09549+0.03323\sqrt{\de}$  \\
\hline 
BH-Ic & $2\ka$   & $\la_0$ & $-2\la_0$  &  $2$ & $\frac{1}{2} \, \de$ & $\frac{2}{\de}$ & $\frac{ \la_0 }{2} \, \de$   \\
CH-Ic & $2\ka$   & $0.2$ & $-0.25$  &  $3.073$ & $0.1742+0.2380\sqrt{\de}$ & $\frac{0.7683}{\de}+\frac{1.050}{\sqrt{\de}}$ & $0.08396+0.04761 \sqrt{\de}$   \\
\hline 
BH-Id & $3\ka$   & $0.2055$ & $-0.3813$  & $2.168$ & $0.4878\,\de$ & $\frac{1.419}{\de}$ & $0.1112\,\de$ \\ 
CH-Id & $3\ka$   & $0.2$ & $-0.25$  &  $2.880$ & $0.1324+0.2289\sqrt{\de}$ & $\frac{0.8226}{\de} + \frac{1.091}{\sqrt{\de}} $ & $0.06732+0.05483\sqrt{\de}$   \\ 
\hline
\end{tabular}
\label{solcase1}
\end{table*}

\begin{table*}
\caption{Representative solutions in the second family ($R_{rr}=0$). We have
taken $\mu_0=0$ and set $\mu_1$ as the length unit. In the columns of ``Event
horizon behavior'', the leading terms of the expansions in Eq.~\rf{hexp} are
given. }
\renewcommand{\arraystretch}{2}
\begin{tabular}{m{1.3cm}m{1.0cm}m{1.3cm}m{1.3cm}m{2.0cm}m{1.4cm}m{3.1cm}m{2.2cm}m{3cm}}
\hline\hline
 Solution & $\xi$ & \multicolumn{3}{l}{Asymptotic parameters} & \multicolumn{4}{c}{Event horizon behavior} \\
\hline
 &  & $\mu_2$ & $\la_0$ & $\la_1$ & $r_h$ & $-g_{tt}$ & $g_{rr}$ & $b_t$  \\
\hline
BH-IIa & $-\ka$ & $0.9$ & $0.2$ & $-0.25$  & $1.680$ & $0.1602 \, \de$ & $\frac{0.7278}{\de}$ & $0.04205+0.07761\,\de$  \\
BH-IIb & $-\ka$ & $1.1$ & $0.2$ & $-0.25$  & $2.222$ & $0.6537 \, \de$ & $\frac{2.699}{\de}$ & $0.09487+0.03065\,\de$  \\
CH-IIa & $-\ka$ & $1.1$ & $0.1$ & $-0.25$  &  $2.603$ & $0.08048+0.2286\sqrt{\de}$ & $\frac{0.5266}{\de}+\frac{1.351}{\sqrt{\de}}$ & $0.001186+0.01286\sqrt{\de}$  \\
\hline 
BH-IIc & $\ka$  & $0.9$ & $0.2$ & $-0.25$  & $1.922$ & $0.4691 \, \de$ & $\frac{1.841}{\de}$ & $0.07564+0.05401\, \de$  \\
BH-IId & $\ka$ & $1.1$ & $0$ & $-0.25$  & $2.129$ & $0.5445 \, \de$ & $\frac{2.279}{\de}$ & $-0.1167+0.04932\,\de$  \\
CH-IIb & $\ka$  & $1.1$ & $0.3$ & $-0.25$  &  $3.674$ & $0.4059+0.04598\sqrt{\de}$ & $\frac{0.003473}{\de}+\frac{0.2002}{\sqrt{\de}}$ & $0.2223+0.01071\sqrt{\de}$  \\
\hline 
BH-IIe & $2\ka$  & $0.9$ & $0.2$ & $-0.25$  & $1.925$ & $0.4886 \, \de$ & $\frac{1.900}{\de}$ & $0.08042+0.05239 \, \de$   \\
BH-IIf & $2\ka$ & $1.1$ & $0.2$ & $-0.25$  & $2.109$ & $0.5100 \, \de$ & $\frac{2.152}{\de}$ & $0.06976+0.07281\,\de$  \\
\hline 
BH-IIg & $3\ka$ & $0.9$  & $0.2$ & $-0.25$  & $1.910$ & $0.4865\,\de$ & $\frac{1.883}{\de}$ & $0.08766+0.04201\,\de$ \\ 
BH-IIh & $3\ka$ & $1.1$ & $0.2$ & $-0.25$  & $2.139$ & $0.5315 \, \de$ & $\frac{2.240}{\de}$ & $0.06163+0.09186\,\de$  \\
CH-IIc & $3\ka$ & $1.1$ & $0.3$ & $-0.5$  &  $2.392$ & $0.03228+0.1915\sqrt{\de}$ & $\frac{1.001}{\de} + \frac{0.3666}{\sqrt{\de}} $ & $0.03170+0.08472\sqrt{\de}$   \\ 
\hline
\end{tabular}
\label{solcase2}
\end{table*}

\begin{figure}
 \includegraphics[width=0.9\linewidth]{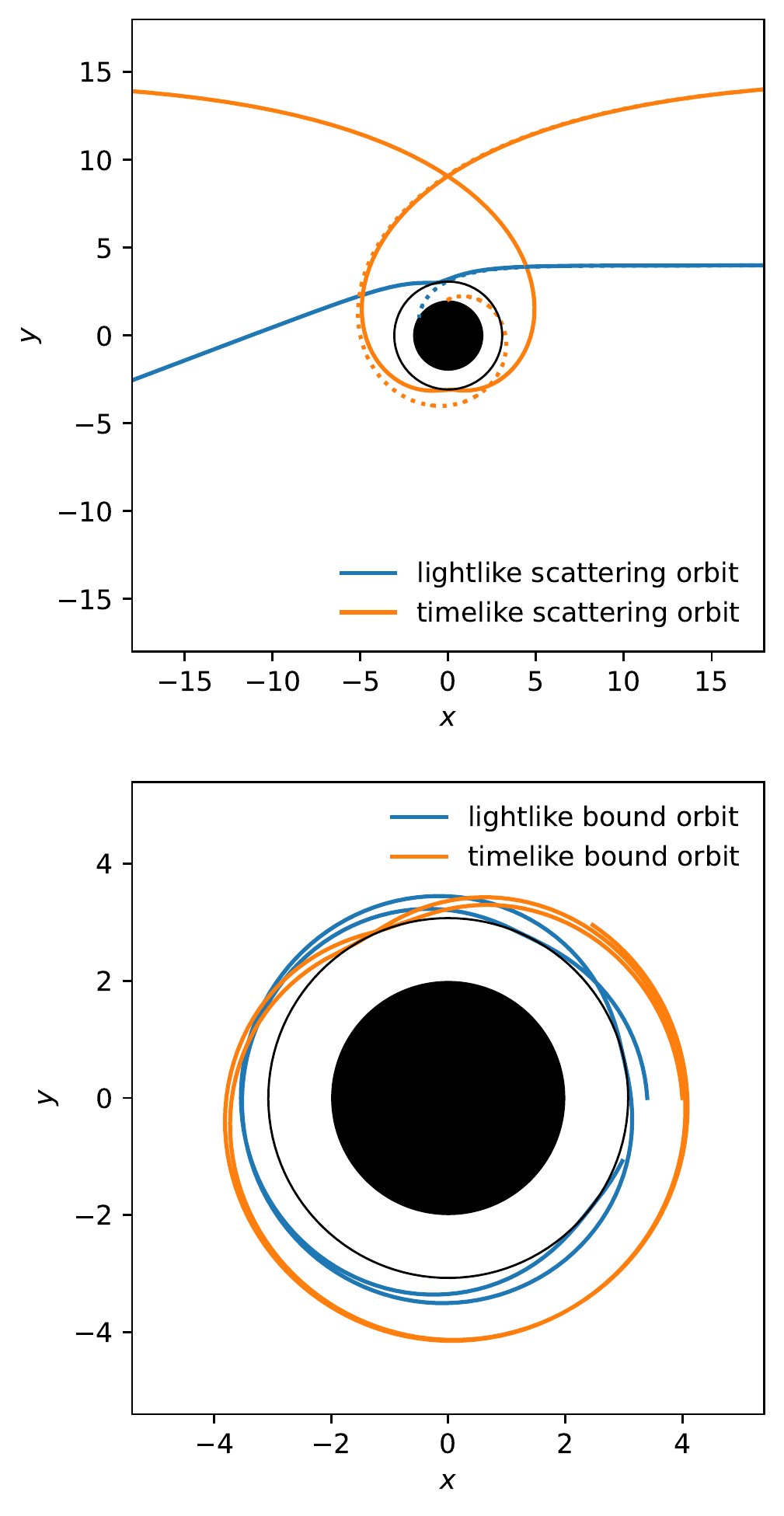}
 \caption{Geodesics bounce back on the event horizon of a CH; the solution
 ``CH-Ic'' in Table~\ref{solcase1} is used. The ADM mass of the solution,
 $\mu_1$, is used as the length unit. The black circles with a radius $r_h \approx
 3.073$ in both panels indicate the event horizon of the CH. The Schwarzschild
 radius, $r_h=2$ is indicated by the solid discs for comparison. In the upper
 panel, the dotted trajectories are the orbits with the same incident velocities
 in the Schwarzschild spacetime that fall into the Schwarzschild BH. }
\label{fig7}
\end{figure}

\subsection{The name ``compact hill'' and its geodesics}
\label{consequences}

The CH solutions share the same form of the asymptotic expansions with the BH
solutions, so they mimic the BH solutions at large radius. But near the event
horizon, the nonvanishing $g_{tt}$ causes a distinctive feature for them. For
CHs, geodesics bounce back on the event horizon, which is the reason for us to
name them ``compact hill''. The very terminology ``event horizon" actually seems
improper as this feature of the CH solutions is radically different from the
event horizon of BHs. However, in the sense that the null hypersurface $r=r_h$
of the CHs prevents geodesics outside of it from going in while the event
horizon of BHs prevents geodesics inside from going out, we still refer $r=r_h$
as the {\it{event horizon}} of the CHs.

Now to show that geodesics bounce back on the event horizon of the CHs, one
first finds that under the metric in Eq.~\rf{ssmetric} the radial components of
the four-velocity and the four-acceleration can be written as 
\bea
&& \left( \frac{dr}{d\la} \right)^2 = e^{-2\mu} \left( \ep^2 \, e^{-2\nu} - \frac{l^2}{r^2} + k \right),
\nonumber \\
&& \frac{d^2r}{d\la^2} = e^{-2\mu} \left[ -\ep^2 \, e^{-2\nu}(\mu'+\nu') + \frac{l^2}{r^2} \left( \mu' + \frac{1}{r} \right) - k \mu' \right] ,
\label{radialgeo}
\eea
where the prime denotes the derivative with respect to $r$, and the integral
constants $\ep$ and $l$ are
\bea
\ep &=& e^{2\nu} \frac{dt}{d\la}, \nonumber \\
l &=& r^2 \frac{d\ph}{d\la}, 
\label{epl}
\eea
while the integral constant $k$ takes $-1,\,  0$, and $1$, for timelike,
lightlike, and spacelike geodesics as $\la$ is the suitable affine parameter in
each case. With $g_{tt} = -e^{2\nu}$ being nonzero on the event horizon, we
immediately see $dr/d\la \rightarrow 0$ on the event horizon. Moreover, using
the expansions in Eq.~\rf{hexp} we find near the event horizon
\bea
&& \left( \frac{dr}{d\la} \right)^2 \rightarrow \frac{r-r_h}{M_{10}} \left( \frac{\ep^2}{N_{10}} - \frac{l^2}{r_h^2} + k \right) ,
\nonumber \\
&& \frac{d^2r}{d\la^2} \rightarrow \frac{1}{2M_{10}} \left( \frac{\ep^2}{N_{10}} - \frac{l^2}{r_h^2} + k \right) . 
\eea 
Because $M_{10}$ is positive, the combination in the parentheses must be
nonnegative for $\left( dr/d\la \right)^2 \ge 0 $. Therefore near the event
horizon the radial acceleration $d^2r/d\la^2$ is nonnegative. Together with the
fact that the radial velocity $dr/d\la$ approaches zero on the event horizon, we
conclude that geodesics bounce back there.

The gravitational force of a CH exerting on a test mass then can be thought as
repulsive near the event horizon while attractive far away. As far as we know
there is completely no evidence for this kind of gravitational interaction in
experiments and astrophysical observations. Anyway, the effect is academically
interesting. The simplest scenario is that a test mass is released at rest and
then oscillates between the initial position and the event horizon. More
generally, bound orbits exist near the event horizon. The repulsive nature near
the event horizon ensures that even the lightlike circular orbit at the radius
determined by $\nu' = 1/r$ is stable. Figure~\ref{fig7} shows example orbits
bouncing back on the event horizon of a CH.

Classical trajectories bounce back on the event horizon, suggesting that GWs
echo, which is a genuine effect that can be searched for in GW observations
\cite{Westerweck:2017hus, Testa:2018bzd}. It turns out that the potentials for
waves are finite on the event horizon so waves are not perfectly reflected. What
marvelous is that the potentials become complex inside the event horizon, and it
is well known in quantum mechanics that complex potentials cause absorption of
waves (e.g. see Refs.~\cite{Avishai:1976bk, Vibok:1992}).  Before we
approximately demonstrate the reflection on the event horizon of a CH using a
scalar wave, let us admit that not only the potentials for waves but also the
Ricci scalar becomes complex inside the event horizon of  CHs. The origination
of their imaginary parts can be seen from Eq.~\rf{hexp}, where $\sqrt{\de}$
becomes imaginary when $r<r_h$. Though it first appears to us that the CH
solutions are unphysical due to the complex Ricci scalar, a second thought about
the absorption property related to the imaginary parts of the potentials strikes
us the idea that a spacetime region with a complex Ricci scalar might be where
information is lost. The idea needs to be further explored, but that is beyond
the scope of the present work.  In the remaining of this section, we only try to
use a brute but manageable approximation to illustrate the reflection of a
scalar wave on the event horizon of a CH.   

Starting with the Klein-Gordon equation
\bea
0 = \Box_g \Ph = \frac{1}{\sqrt{-g}} \prt_\mu \left( g^\mn \sqrt{-g}\, \prt_\nu \Ph \right) ,
\eea
and using the spherical metric in Eq.~\rf{ssmetric} as well as the spherical
wave ansatz for the scalar field $\Ph = e^{i\si t} \, \Ps(r)$, one obtains an
equation for $\Ps(r)$
\bea
\Ps'' + \left( \nu' - \mu' + \frac{2}{r} \right) \Ps' + \si^2 e^{2(\mu-\nu)} \Ps = 0 .
\label{radialeq1}
\eea
The changes of variables 
\bea
z &=& r \Ps, \nonumber \\
\frac{dr_{*}}{dr} &=& e^{\mu-\nu}
\eea
put Eq.~\rf{radialeq1} into the standard form \cite{Berti:2009kk}
\bea
\frac{d^2z}{dr_*^2} + \left( \si^2 - V_s \right) z = 0,
\label{radialeq2}
\eea
with the potential for the scalar wave being 
\bea
V_s =  e^{2(\nu-\mu)} \frac{\nu'-\mu'}{r} .
\eea   
Fully solving the scattering problem of Eq.~\rf{radialeq2} is beyond the scope
of the present work. We shall focus on the behavior of the wave near the event
horizon, where the potential $V$ and the derivative $dr_*/dr$ have the
expansions
\bea
&& V_s = \frac{1}{2r_h} \frac{N_{10}}{M_{10}} \left[ 1 + \frac{3}{2} \left( \frac{N_{20}}{N_{10}} - \frac{M_{20}}{M_{10}} \right) \sqrt{\de} + ... \right] , 
\nonumber \\
&& \frac{dr_*}{dr} = \sqrt{ \frac{M_{10}}{N_{10}} } \frac{1}{\sqrt{\de}} \left[ 1 - \frac{1}{2} \left( \frac{N_{20}}{N_{10}} - \frac{M_{20}}{M_{10}} \right) \sqrt{\de} + ... \right] .
\label{radialeq3}
\eea
Besides being complex inside the event horizon for the potential $V_s$,
Eq.~\rf{radialeq3} is different from the BH expressions because of two other
features. First, for BHs $V_s$ is zero on the event horizon while
Eq.~\rf{radialeq3} is not. Then, the tortoise radius $r_*$ goes to negative
infinity on the event horizon of BHs, but now that $dr_*/dr$ is only divergent
as $1/\sqrt{\de}$ for the CHs we find $r_*$ proportional to $\sqrt{\de}$ plus an
integral constant.

Taking the distinctive features of Eq.~\rf{radialeq3} into consideration, we
reckon that the step potential
\bea
V_{\rm step} = 
\begin{cases}
V_{s1} , & {\rm Re}(r_*) \ge 0 \\
V_{s2} + i \Ga_2 , & {\rm Re}(r_*) < 0
\end{cases}
\label{stepv}
\eea
might be used to approximate the effective potential $V_s$ near the event
horizon. In Eq.~\rf{stepv}, the real constant $V_{s1}$ and the complex constant
$V_{s2} + i \Ga_2$ can be thought as certain average values of $V_s$ right
outside and inside the event horizon. The tortoise radius, which also becomes
complex inside the event horizon, has been set to zero at the event horizon.
With $V_s$ brutely approximated by $V_{\rm step}$, the scattering problem of
Eq.~\rf{radialeq2} is analytically solvable. Assuming no waves coming out of the
center at $r=0$, the plane wave ansatz 
\bea
z = 
\begin{cases}
C_+ e^{ik_1 r_*} + C_- e^{-ik_1 r_*}  , & {\rm Re}(r_*) \ge 0 \\
D_+ e^{ik_2 r_*} , & {\rm Re}(r_*) < 0
\end{cases}
\label{planez}
\eea
satisfies the equation when
\bea
k_1 &=& \sqrt{ \si^2 - V_{s1} }, \nonumber \\
k_2 &=& \sqrt{ \si^2 - V_{s2} - i \Ga_2 }.
\eea  
The continuity of $z$ and $dz/dr_*$ at $r_* = 0$ fixes the ratio between the
amplitudes of the reflected and the incident waves, which is
\bea
\frac{C_-}{C_+} = \frac{k_1-k_2}{k_1 + k_2}.
\eea
The reflection rate is then
\bea
{\cal{R}} = \bigg| \frac{C_-}{C_+} \bigg|^2 = \frac{ k_1^2 + |k_2|^2 - 2k_1 {\rm Re}(k_2) }{ k_1^2 + |k_2|^2 + 2k_1 {\rm Re}(k_2) } ,
\label{reflrate}
\eea
where we have assumed the energy of the incident wave $\si^2$ to be greater than
$V_{s1}$ so that $k_1$ is real. If we define $\th_2 \in [-\pi, \pi)$ by
\bea
&& \cos\th_2 = \frac{\si^2-V_{s2}}{\sqrt{\left(\si^2-V_{s2}\right)^2 + \Ga_2^2}}, 
\nonumber \\
&& \sin\th_2 = \frac{-\Ga_2}{\sqrt{\left(\si^2-V_{s2}\right)^2 + \Ga_2^2}},
\eea 
then $k_2$ can be explicitly written as 
\bea
k_2 = \left[ \left(\si^2-V_{s2}\right)^2 + \Ga_2^2 \right]^{ \frac{1}{4} } e^{\frac{i\th_2}{2}} .
\eea

A brief understanding on how $\Ga_2$ affects the reflection rate can be achieved
by assuming a small $\Ga_2$ so that Eq.~\rf{reflrate}, up to the leading order
of $\Ga_2$, reads
\bea
{\cal{R}} \approx \left( \frac{k_1-\tilde k_2}{k_1+\tilde k_2} \right)^2 + \frac{ k_1 \left( 3 \tilde k_2^2 - k_1^2\right)}{2 \, \tilde k_2^3 (k_1+\tilde k_2)^4 } \Ga_2^2,
\eea
where $\tilde k_2 = \sqrt{\si^2 - V_{s2}}$. The first term independent of
$\Ga_2$ is the usual reflection rate due to a real step potential. The second
term proportional to the square of $\Ga_2$ represents the contribution from the
imaginary part of the potential. The contribution is positive when $3 \tilde
k_2^2 - k_1^2 > 0$, namely
\bea
\si^2 > \frac{3V_{s2}-V_{s1}}{2} .
\label{inccond}
\eea 
Because we have required $\si^2 > V_{s1}$, Eq.~\rf{inccond} is always satisfied
if $V_{s1} \ge V_{s2}$. If $V_{s1} < V_{s2}$, then the imaginary potential
contributes negatively to the reflection rate when the frequency of the incident
wave is lower than the value in Eq.~\rf{inccond}.

\section{Observational tests}
\label{sec:III}
The exotic solutions of CHs are not supported by current observations in the strong-field regime such as the GW detections and the imagies of the supermassive black holes. We consider testing the BH solutions in the bumblebee model against observations in this section.

\subsection{Classical tests}
\label{sec:IIIa}

Classical tests of the weak-field metric in the bumblebee model using
observations in the Solar System was first done by \citet{Hellings:1973zz}
within the PPN framework. It is equivalent to testing the metric asymptotic
expansion in Eq.~\rf{asyexp} with $\mu_0=0$. In Appendix~\ref{app5}, we
transform the metric expansion in Eq.~\rf{asyexp} to the harmonic coordinates so
that the PPN parameters $\be$ and $\ga$ \cite{Poisson:2014} can be directly read
off. Existing constraints on the PPN parameters can thus be translated to
constrain the free coefficients in the asymptotic expansions in Eq.~\rf{asyexp}.
This is an easy way to obtain constraints on the free parameters in our
solutions. But to clearly see how the parameters in our solutions enter the classical effects and thus obtaining tailored constraints for them, we reckon that it is worth leaving the convenient PPN framework but
calculating predictions directly using Eq.~\rf{asyexp} to compare with
observations.

Considering three classical tests of GR, the advance of perihelion, the
deflection of light, and the Shapiro time delay, we find that the effect of the
advance of perihelion is sensitive to the coefficients $\mu_0,\, \mu_1,\, \nu_1$
and $\nu_2$ while the effects of the other two are only sensitive to $\mu_0, \,
\mu_1$ and $\nu_1$.\footnote{The other classical test of GR, the gravitational
redshift, only tests the equivalence principle~\cite{Will:2018bme}, and it is
left out of the current study because the bumblebee gravity model has already
satisfied the Einstein equivalence principle.} The calculations are done in the
weak-field regime where the ratio of the central mass to the typical length
involved is small so that the results are kept to the linear order of this
ratio. The results for the deflection of light and the Shapiro time delay are
the same as those derived in Ref.~\cite{Casana:2017jkc}, where tight constraints
have been set on the corresponding parameter $l=e^{2\mu_0}-1$. We focus on the
new modification introduced by the coefficient $\nu_2$ in the effect of the
advance of perihelion. 

For a bound timelike geodesic with perihelion separation $r_{\rm min}$ and
aphelion separation $r_{\rm max}$, the advance of perihelion per orbit is given
by
\bea
\de \ph = 2\int_{r_{\rm min}}^{r_{\rm max}} \Big| \frac{d\ph}{dr} \Big| \, dr - 2\pi .
\eea
Using Eqs.~\rf{radialgeo} and \rf{epl} to calculate $dr/d\ph$ and substituting
the expansions for $\mu$ and $\nu$ in Eq.~\rf{asyexp} into it, up to the leading
order of $\mu_1/r_{\rm min}$, the advance of perihelion turns out to be
\bea
\de \ph \approx 2\pi\left( e^{\mu_0} - 1 \right) + \pi e^{\mu_0} \left( \mu_1 - \nu_1 + \frac{\nu_2}{ \nu_1} \right) \frac{ r_{\rm min} + r_{\rm max} }{r_{\rm min} r_{\rm max}} .
\label{aop}
\eea 
Because there are two families of solutions with different recurrence relations
for the coefficients in the asymptotic expansions, we discuss them separately.

For the family of solutions with $b_r=0$, the first few recurrence relations are
given in Eq.~\rf{case1asyrec} with three free parameters, $\mu_1, \, \la_0$ and
$\la_1$. Using the recurrence relations, Eq.~\rf{aop} becomes
\bea
\de \ph &\approx& \pi \left( 3 - \frac{ 2\xi \la_0^2 + 2\xi \la_0 \tilde \la_1 + (\xi-\ka) \tilde \la_1^2 }{ 4\xi \left( 1 - \frac{\xi}{2\ka} \right) \la_0^2 - 4 } \right)
\nonumber \\
&& \times \frac{ \mu_1 \left( r_{\rm min} + r_{\rm max} \right) }{r_{\rm min} r_{\rm max}} ,
\label{aop1}
\eea 
where $\tilde \la_1 = \la_1/\mu_1$.  The recurrence relations in
Eq.~\rf{case1asyrec} are for both BH solutions and CH solutions. If we restrict
our attention to the BH solutions in this family, we should keep in mind that
there is only one free parameter among $\la_0$ and $\la_1$. It is clear that
Eq.~\rf{aop1} reproduces the well-known result
\bea
\de \ph_{\rm GR} &\approx& \frac{ 3 \pi \mu_1 \left( r_{\rm min} + r_{\rm max} \right) }{r_{\rm min} r_{\rm max}} 
\eea
for the Schwarzschild metric when $\la_0=\la_1=0$. In addition, the bumblebee
model recovers the Einstein-Maxwell theory when $\xi=0$, so Eq.~\rf{aop1} gives
the result for the Reissner-Nordstr\"om metric when $\xi=0$ with $\la_1$ set to
$Q/\sqrt{\ka/2}$.

For the family of solutions with $R_{rr}=0$, the first few recurrence relations
are given in Eq.~\rf{case2asyrec} with 5 free parameters, $\mu_0, \, \mu_1,
\mu_2\, \, \la_0$ and $\la_1$.  Using the recurrence relations, Eq.~\rf{aop}
becomes
\bea
\de \ph \approx 2\pi\left( e^{\mu_0} - 1 \right) + e^{\mu_0} \left( 7 + 2 \tilde \mu_2 \right) \frac{ \pi \mu_1 \left( r_{\rm min} + r_{\rm max} \right) }{ 3 r_{\rm min} r_{\rm max}} ,
\label{aop2}
\eea 
where $\tilde \mu_2 = \mu_2/\mu_1^2$. We see that the bumblebee parameters
$\la_0$ and $\la_1$ do not affect the advance of perihelion at the leading order
of $\mu_1/r_{\rm min}$ in this family. Setting $\tilde \mu_2 = 1$, Eq.~\rf{aop2}
reproduces the result in Ref.~\cite{Casana:2017jkc} which in turn goes back to
the Schwarzschild result if taking $\mu_0=0$.

\begin{table*}
\caption{Orbital parameters of 6 Solar-System planets and the fitting residuals
of their perihelion advances.\footnote{Orbital parameters are taken from
\url{https://nssdc.gsfc.nasa.gov/planetary/factsheet/}.} }
\begin{tabular}{m{1.7cm} m{3.2cm} m{3.2cm} m{2.7cm} m{5.5cm}}
\hline \hline
Planet & Perihelion ($10^6 \, {\rm km}$) & Aphelion ($10^6 \, {\rm km}$) & Period (days) & Perihelion advance residual (mas\,yr$^{-1}$) \\
\hline
Mercury  &  46.00 & 69.82 & 87.97 & $-0.020\pm 0.030$   \\
Venus  &  107.5 & 108.9 & 224.7 & $0.026\pm 0.016$  \\
Earth  &  147.1 & 152.1 & 365.2 & $0.0019\pm 0.0019$ \\
Mars  &  206.7 & 249.3 & 687.0 & $-0.00020\pm 0.00037$  \\
Jupiter  & 740.6 & 816.4 & 4331 & $0.59\pm 0.28$  \\
Saturn  &  1358 & 1507 & 10747 & $-0.0032\pm 0.0047$ \\
\hline
\end{tabular}
\label{perihelionprecession}
\end{table*}

\begin{figure*}
 \includegraphics[width=\linewidth]{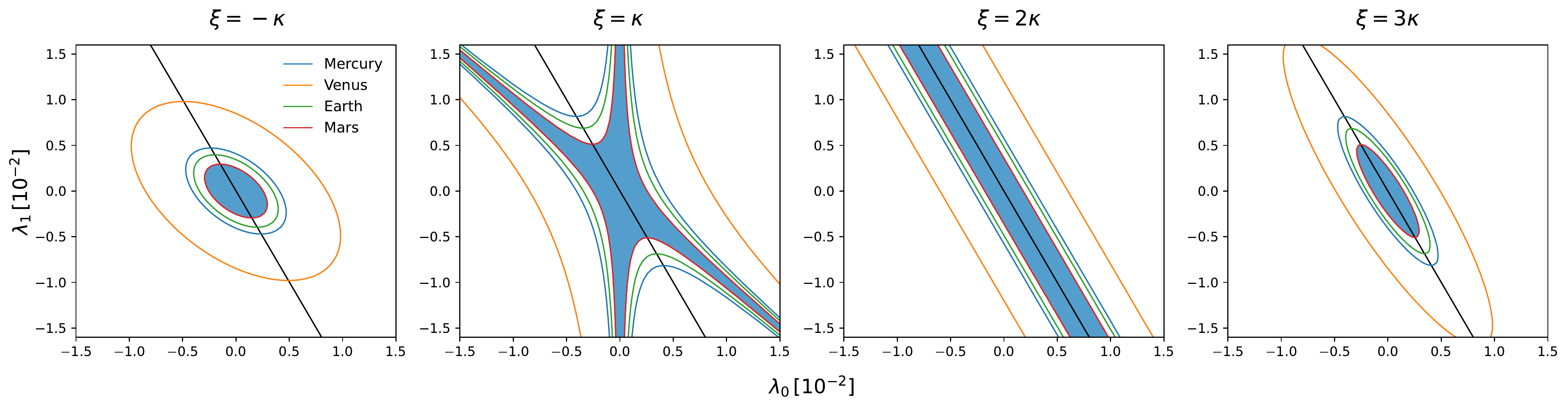}
 \caption{Constraints on the bumblebee parameters $\la_0$ and $\la_1$ for the
 first family of solutions ($b_r=0$) from the fitting residuals of perihelion
 advances in Table~\ref{perihelionprecession}. The metric parameter $\mu_1$ is
 used as the unit. In these plots, BHs exist along the black lines passing
 through the origin in the $\la_0$-$\la_1$ plane (the brown lines in
 Fig.~\ref{fig3}). Note that Jupiter and Saturn give constraints too loose to
 shown in the plots and they are omitted. }
\label{fig8}
\end{figure*}
\begin{figure}
 \includegraphics[width=0.95\linewidth]{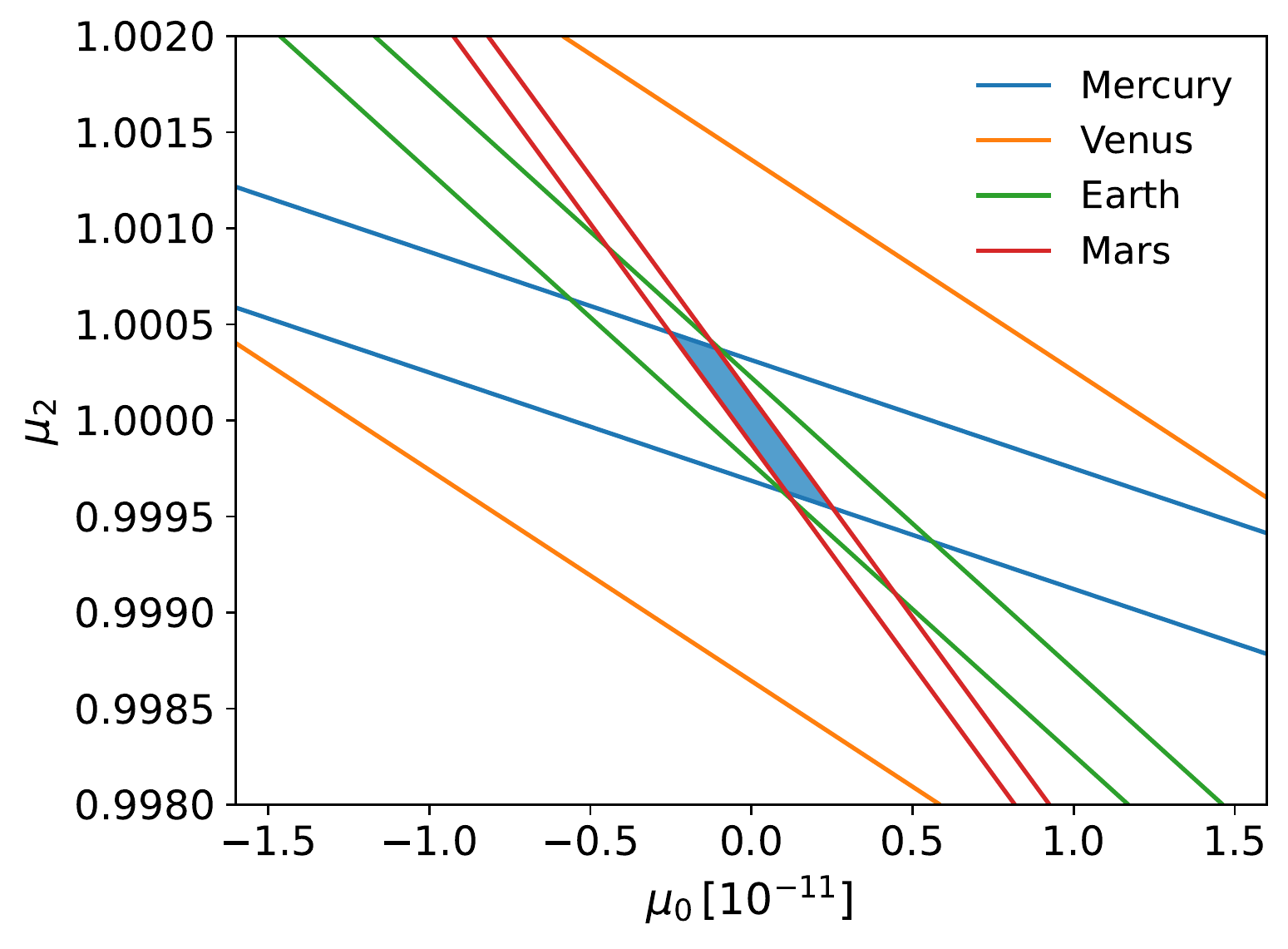}
 \caption{Constraints on the metric parameters $\mu_0$ and $\mu_2$ for the
 second family of solutions ($R_{rr}=0$) from the fitting residuals of
 perihelion advances in Table~\ref{perihelionprecession}. The metric parameter
 $\mu_1$ is used as the unit. Note that Jupiter and Saturn give constraints too
 loose to shown here and they are omitted. }
\label{fig9}
\end{figure}

\citet{Pitjeva:2013xxa} fitted the advances of perihelia of 6 planets using
standard gravity described by GR. We collect their fitting residuals as well as
the orbital parameters in Table~\ref{perihelionprecession}. We use the larger
between a fitting residual and its uncertainty as an upper bound for $|\de
\ph-\de \ph_{\rm GR}|$. Then constraints on $\la_0$ and $\la_1$ for the first
family of solutions can be set using Eq.~\rf{aop1}, and constraints on $\mu_0$
and $\mu_2$ can be set for the second family of solutions using Eq.~\rf{aop2}.
The results are shown in Figs.~\ref{fig8} and \ref{fig9}. As expected,
parameters in both solution families are severely constrained so that the metric
outside the Sun is reasonably the Schwarzschild solution to a good
approximation. Notice that the constraints on $\mu_0$ in Fig.~\ref{fig9} are
consistent with the results in Ref.~\cite{Casana:2017jkc}. In the future, a
global fit of data directly with expressions given by the bumblebee gravity
model is desirable to further eliminate degeneracy among orbital parameters and
theory parameters.

Though both pointing to the Schwarzschild solution, we feel that the constraints
for the two families of solutions have different interpretations. For the first
family of solutions, the constraints are on the bumblebee parameters $\la_0$ and
$\la_1$, meaning that the bumblebee charge of the Sun must be very small if not
vanishing. Naively, different objects can carry different amount of bumblebee
charge, so these constraints are not necessarily valid for other types of
astrophysical objects, like BHs or neutron stars. For the second family of
solutions, the constraints are directly on the metric parameters $\mu_0$ and
$\mu_2$ regardless of the bumblebee parameters $\la_0$ and $\la_1$. So they are
supposed to be general and valid for all types of astrophysical objects.

\subsection{BH images}  
\label{sec:IIIb}  

Advances of perihelia of the planets in the Solar System set stringent
constraints on the parameters of the static spherical solutions in the bumblebee
gravity model. But the results are doubtfully universal for the first family of
solutions where the constraints are on the bumblebee parameters $\la_0$ and
$\la_1$. While the Sun has tiny or zero bumblebee charge, more compact objects
like BHs may still possess a considerable amount of it. So we consider to put
bounds on the bumblebee charge carried by the supermassive BHs whose shadow
images have been recently taken by the EHT
Collaboration~\cite{EventHorizonTelescope:2019dse,
EventHorizonTelescope:2022xnr}. 

The mathematical shape of the shadow of a BH depends on the mass and the spin of
it. When extracting from observations, the precise shape is also influenced
by the observing resolution as well as the structure of the surrounding matter
that emits EM waves. The mass of the BH determines the size of the shadow, while
other elements contribute corrections typically of order unity
\cite{EventHorizonTelescope:2019dse}. For our purpose of setting preliminary
bounds on the bumblebee charge, it is sufficient to consider the simplest model
where a circular shadow is created by a static spherical BH. In this case, the
radius of the shadow is the impact parameter of a lightlike geodesic infinitely
approaching the lightlike circular orbit, i.e., the light ring. In the static
spherical spacetime described by Eq.~\rf{ssmetric}, the light ring, if exists,
is at a radius $r_{\rm lr}$ given by 
\bea
\nu'\big|_{r=r_{\rm lr}} = \frac{1}{r_{\rm lr}} ,
\eea
and the limit of the impact parameter corresponding to the light ring is 
\bea
\si_{\rm lr} = r e^{-\nu} \, \big|_{r=r_{\rm lr}} .
\label{ssshadow}
\eea

Denoting the distance from the observer to the BH as $D$, the angular diameter
of the shadow is
\bea
d = \frac{2\si_{\rm lr}}{D} = 2 \th_g \frac{\si_{\rm lr}}{M}, 
\eea  
where $M=\mu_1$ is the mass of the BH and $\th_g = M/D$. Adopting the
observational results 
\bea
d &=& 42 \pm 3 \, \mu {\rm as}, \nonumber \\
\th_g &=& 3.62 \pm 0.60 \, \mu {\rm as}
\eea
according to Refs.~\cite{Gebhardt:2011yw, EventHorizonTelescope:2019dse} for the
supermassive BH in M87, we get 
\bea
\frac{\si_{\rm lr}}{M} = 5.80 \pm 1.05 .
\label{m87}
\eea
Adopting the observational results\footnote{The average of the results from the
Very Large Telescope Interferometer (VLTI) and Keck is used as the value for
$\th_g$, and their difference is used as the uncertainty.} 
\bea
d &=& 51.8 \pm 2.3 \, \mu {\rm as}, \nonumber \\
\th_g &=& 5.02 \pm 0.20 \, \mu {\rm as}
\eea
according to Refs.~\cite{Do:2019txf, GRAVITY:2021xju,
EventHorizonTelescope:2022xnr} for the supermassive BH in the Milky Way, we get 
\bea
\frac{\si_{\rm lr}}{M} = 5.16 \pm 0.31 .
\label{sgra}
\eea

\begin{figure}
 \includegraphics[width=0.95\linewidth]{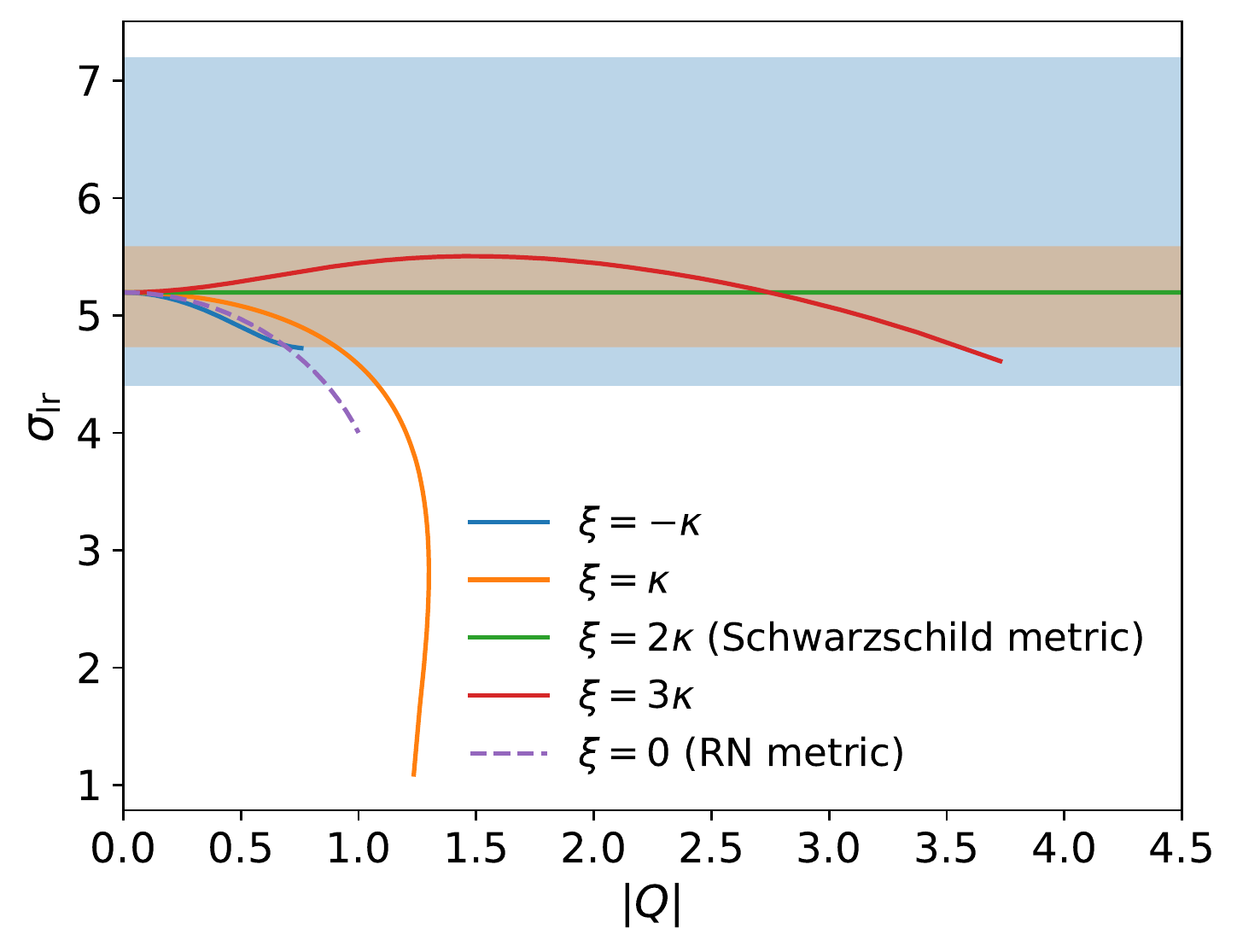}
 \caption{Radius of the shadow versus the bumblebee charge for static spherical
 BHs in the bumblebee gravity model. The two colored bands indicate the
 observational results in Eqs.~\rf{m87} and \rf{sgra}, with the orange band for
 Sgr A$^*$ and the blue band for M87$^*$. The mass of the BH is used as the
 unit. }
\label{fig10}
\end{figure}

In Fig.~\ref{fig10}, numerically calculated $\si_{\rm lr}$ using
Eq.~\rf{ssshadow} for the solutions in Fig.~\ref{fig5} is plotted with respect
to the bumblebee charge. The observational results in Eqs.~\rf{m87} and
\rf{sgra} are indicated by the shaded bands. We see that for different coupling
constant $\xi$, the bounds on the bumblebee charge can be very different. For example, if we look at the bounds from the result of Sgr A$^*$, the bumblebee charge is up to about 0.7 when $\xi=-\ka$ but to about $3.5$ when $\xi=3\ka$.

\section{Summary}
\label{sec:sum}

The bumblebee gravity model is an important vector-tensor gravity theory with
Lorentz-symmetry violation.  Our systematic study of the  static spherical
vacuum solutions in the bumblebee gravity model reveals two families of
solutions that have diverging $g_{rr}$ but finite curvature scalars at a finite
radius $r_h$. The first family, characterized by a bumblebee background field
with only the temporal component, consists of solutions given by three free
parameters. The second family, characterized by a vanishing $rr$-component of
the Ricci tensor, consists of solutions given by 5 free parameters. In each
family, BH solutions exist under the requirement that $g_{tt}$ vanishes at the
event horizon $r_h$. The remaining solutions that have nonzero $g_{tt}$ at $r_h$
are dubbed ``CHs'' for geodesics bounce back at the event horizon (see
Sec.~\ref{consequences}).

The fascinating CH solutions provide a gravitational interaction that attracts
in the same way as the usual gravity far away from the center but is strongly
repulsive near the event horizon. Consequently, unlike BHs, stable bound orbits
exist right outside the event horizon of CHs. If waves are considered instead of
classical geodesics, one finds that the potentials for waves are finite at the
event horizon of CHs so that waves can transit into the event horizon while
being partially reflected. Furthermore, because the potentials are complex
inside the event horizon of CHs, the waves entering the event horizon will be at
least partially absorbed by spacetime itself, causing possible loss of
information.   We concede that no experiments and observations support any of
these features. However, the echoes of waves at the event horizon of CHs and the
loss of information associated with spacetime that has complex potentials for
waves and complex Ricci scalar are certainly interesting issues worth further
study.

Restricting our attention to the BH solutions, we find that in the first family,
the number of free parameters reduces from three to two, namely the mass and the
charge, while in the second family there are still 5 free parameters. The BH
solutions in both families have their merits. In the first family, the solutions
recover the Reissner-Nordstr\"om metric when $\xi=0$. In this sense, they are
the expected BH solutions when the coupling between the Ricci tensor and the
vector field is added to the Einstein-Maxwell theory. In the second family,
there is an analytical solution given by Eqs.~\rf{asol1} and \rf{asol2}, which
deserves a mention because by setting $\mu_0=0$ the Schwarzschild metric is
recovered together with a nontrivial bumblebee background field.  It is amusing
to notice the special case of $\xi=2\ka$, for both families give the
Schwarzschild metric together with a bumblebee background field that has only a
temporal component proportional to $1-2M/r$.  

In Sec.~\ref{sec:III}, the BH solutions are tested using Solar-System
observations and the images of the supermassive BHs. For the solutions in the
first family, constraints on the bumblebee charge carried by the central objects
are obtained. The upper bound for the ratio of the bumblebee charge to the mass is at the order of
$10^{-3}$ for the Sun while it roughly varies from $0.7$ to $3.5$ depending on
the value of $\xi$ for the supermassive BHs. We need to point out that the
bumblebee charge is completely unconstrained when $\xi=2\ka$ as the metric is
exactly Schwarzschild in this case. For the solutions in the second family,
stringent constraints on the metric parameters are obtained, suggesting that if
static spherical objects are described by the solutions in this family, the
spacetime is most likely to be Schwarzschild but can be accompanied by a
nontrivial bumblebee background field.

In conclusion, as a generalization of the Einstein-Maxwell theory and an
important example for Lorentz-violating gravitational theories, the bumblebee gravity
model in Eq.~\rf{actionB} indeed contains richer solutions. Testing these
solutions in experiments and observations is undoubtedly a worthy subject for
fundamental physics. On the one hand, the bumblebee background field showing up
from nowhere in the vacuum solutions defines special local frames so that local
Lorentz symmetry is broken. Therefore testing whether there is any such
background field is testing the fundamental symmetry of our Universe. On the
other hand, if the bumblebee field is regarded as the EM vector potential,
testing the solutions in the bumblebee model will tell us whether the EM field
only minimally couples with gravity as in the Einstein-Maxwell theory or the
nonminimal coupling between the Ricci tensor and the vector field exists. The
tests done in this work have not excluded the possibility of a bumblebee
background field though the metric is constrained to be close to the
Schwarzschild metric in the spherically symmetric situation. The next step is
naturally to put the solutions into test against the GW observations. Future
works on extending the spherical BH solutions to rotating BHs, calculating the
quasinormal modes of the BHs, as well as solving the two-body problem in the PPN
framework, are directions that need to be explored.

\acknowledgments 

We thank Bin Chen and V. Alan Kosteleck\'y for insightful discussions and comments. We are also grateful to the anonymous referee for helpful feedback.
This work was supported by the National Natural Science Foundation of China (11991053, 12147120, 11975027,
11721303), the China Postdoctoral Science Foundation (2021TQ0018), the National
SKA Program of China (2020SKA0120300), the Max Planck Partner Group Program
funded by the Max Planck Society, and the High-Performance Computing Platform of
Peking University. R.X.\ is supported by the Boya Postdoctoral Fellowship at
Peking University.

\appendix

\section{Field equations under the static spherical ansatz}
\label{app1}

Denoting $E_\mn = G_\mn - \ka \left( T_b \right)_\mn$, with the static spherical
ansatz the nonvanishing Einstein field equations are
\bw
\bea
0 = E_{tt} &=& \frac{e^{2 \nu-2 \mu} \left(1 - 2 r \mu' - e^{2 \mu}\right)}{r^2}  - \xi e^{-2\mu} \left[ b_t^{\prime\prime} - b_t^{\prime} \left( \mu'+ 4  \nu'-\frac{2}{r}\right) + b_t  \left(  2\mu'\nu' - \frac{4}{r} \nu'- 2  \nu''+ \nu'^2 \right) \right] b_t + \left( \frac{\ka}{2} - \xi \right)e^{-2 \mu} b_t^{\prime\,2}
\nonumber \\
&&  + \xi e^{2 \nu-4 \mu} \left[ b_r^{\prime\prime}-b_r^{\prime} \left(5  \mu'-\frac{4}{r}\right) + b_r \left(- \mu''+3 \mu'^2 - \frac{6}{r} \mu'+ \frac{1}{r^2} \right) \right] b_r + \xi e^{2 \nu-4 \mu} b_r^{\prime\,2} ,
\nonumber \\
0 = E_{rr} &=& \frac{e^{2 \mu}-2 r \nu'-1}{r^2} - e^{-2 \nu} \left( \frac{\ka}{2} b_t^{\prime\,2} - \xi b_t b_t^{\prime} \nu' + \xi b_t^2 \nu'^2\right) - \xi e^{-2\mu} \left[ b_r^{\prime} \left( \nu'+\frac{2}{r}\right) - b_r \left( \nu''+ \nu'^2- \frac{2}{r} \nu'-\frac{1}{r^2}\right) \right] b_r ,
\nonumber \\
0 = E_{\th\th} &=& \frac{E_{\ph\ph}}{\sin^2\th} = r^2 e^{-2 \mu} \left( \mu'\nu'+ \frac{1}{r}\mu'-\frac{1}{r}\nu'- \nu''- \nu'^2\right) + r^2 e^{-2 (\mu+\nu)} \left( \frac{\ka}{2} b_t^{\prime\,2} - \xi b_t b_t^{\prime} \nu' + \xi b_t^2 \nu'^2 \right) - \xi r^2 e^{-4\mu} b_r^{\prime\,2}  
\nonumber \\
&& \hskip 1.2cm - \xi r^2 e^{-4 \mu} \left[ b_r^{\prime\prime} - b_r^{\prime} \left(5 \mu'-2 \nu'- \frac{2}{r}\right) + b_r \left(3 \mu' \left( \mu'- \nu'- \frac{1}{r}\right)+ \nu''- \mu''+ \nu'^2+ \frac{\nu'}{r}\right) \right] b_r ,
\nonumber \\
0 = E_{tr} &=& \frac{\xi}{\ka} e^{-2 \mu} \left( \mu' \nu' + \frac{2}{r} \mu' -\nu''-\nu'^2 \right) b_t b_r ,
\label{ssefe}
\eea 
\ew
and the vector field equation gives
\bea
0 &=& b_t^{\prime\prime}- b_t^{\prime} \left( \mu'+ \nu'-\frac{2}{r}\right)+ \frac{\xi}{\ka} b_t \left( \mu' \nu' - \frac{2}{r}\nu' - \nu'' - \nu'^2 \right) ,
\nonumber \\
0 &=& \frac{\xi}{\ka} \left( \mu' \nu' + \frac{2}{r} \mu' -\nu''-\nu'^2 \right) b_r ,
\label{ssvfe} 
\eea 
where the prime indicates the derivative with respect to $r$. Noticing $R_{rr} =
\left( \mu' \nu' + 2\mu'/r -\nu''-\nu'^2 \right)$, the second equation in
Eq.~\rf{ssvfe} leads to two possibilities:
\begin{enumerate} 
\item $b_r = 0$, in which case $\mu$ and $\mu'$ can be eliminated from the
equations, so we arrive at a system of two second-order ODEs for $\nu$ and
$b_t$; 
\item $R_{rr} = 0$, in which case we find that $b_r$ can be solved in terms of
other variables,
\bw
\bea
b_r^2 &=& \frac{ e^{2 \mu-2 \nu} }{3\xi} \frac{ r \left[ (2\ka-\xi)r \nu' + \ka - 2\xi \right] b_t^{\prime\,2} + 6\xi r \nu' b_t b_t' - 2\xi \left[ \left(\frac{\xi}{\ka}-1\right) \left(  r \mu'\nu'+2\mu' + 2\nu' \right) + \left(\frac{\xi}{\ka}+1\right) r\nu^{\prime\,2} \right] b_t^2 }{ r \mu'\nu' + 2\mu' - r \nu'^2 + \nu' } 
\nonumber \\
&& - \frac{e^{2 \mu}}{\xi}  \frac{ r \mu' \nu'+2 \mu' - r \nu^{\prime\,2} +  \nu' + e^{2 \mu} \nu' }{ r \mu'\nu' + 2\mu' - r \nu'^2 + \nu' } ,
\label{case2brsol}
\eea
\ew
so the system consists of three second-order ODEs for $\mu, \, \nu$ and $b_t$.  
\end{enumerate}

\section{Asymptotic expansions of the variables}
\label{app2}

For the case of $b_r=0$, substituting the expansions in Eq.~\rf{asyexp} into
Eqs.~\rf{ssefe} and \rf{ssvfe}, we find the following recurrence relations
\bea
&& \mu_0 = 0, 
\nonumber \\
&& \nu_1 = -\mu_1 , 
\nonumber \\
&& \nu_2 = -\mu_1^2 + \frac{ 2\xi \mu_1^2 \la_0^2 + 2\xi \mu_1 \la_0 \la_1 + (\xi-\ka) \la_1^2 }{ 4 \left[\xi \left( 1 - \frac{\xi}{2\ka} \right) \la_0^2 - 1 \right] } ,
\nonumber \\
&& \mu_2 = \mu_1^2 + \frac{ \xi(1-\frac{\xi}{2\ka}) \la_0^2 \left[2\xi \mu_1^2 \la_0^2 + 2\xi \mu_1 \la_0 \la_1 + \ka \la_1^2 \right] }{ 4\left[ \xi (1 - \frac{\xi}{2\ka}) \la_0^2  - 1\right] } 
\nonumber \\
&& \hskip 0.6cm - \frac{  6\xi\mu_1^2 \la_0^2 + 6\xi \mu_1 \la_0 \la_1 + (2\xi-\ka) \la_1^2 }{ 4\left[\xi (1 - \frac{\xi}{2\ka}) \la_0^2  - 1\right] } ,
\nonumber \\
&& \la_{2} = \frac{ \xi \la_0 \left[2\xi \mu_1^2 \la_0^2 + 2\xi \mu_1 \la_0 \la_1 + (\xi-\ka) \la_1^2 \right]}{ 4\ka \left[\xi \left( 1 - \frac{\xi}{2\ka} \right) \la_0^2 - 1 \right]} ,
\nonumber \\
&& ... \, . 
\label{case1asyrec}
\eea
Since the system of equations can be reduced to two second-order ODEs, there
should be 4 free coefficients in principle. However, we have dropped the
constant term $\nu_0$ in the expansion of $\nu$ due to the freedom to redefine
the time coordinate. There are three independent coefficients unfixed as shown
in Eq.~\rf{case1asyrec}. They are taken as $\mu_1, \, \la_0$ and $\la_1$.

In this family of solutions, the Schwarzschild metric
\bea
\nu = -\mu = \frac{1}{2} \ln{\left(1- \frac{2M}{r}\right)}
\eea
is recovered by taking $\mu_1=M$ and $\la_0=\la_1=0$. We also note that the
Reissner-Nordstr\"om metric
\bea
\nu = -\mu = \frac{1}{2} \ln{\left(1-\frac{2M}{r} + \frac{Q^2}{r^2} \right)}
\label{rnmetric}
\eea
is recovered by taking $\mu_1=M$ and $\la_1^2 = 2Q^2/\ka$ specifically for
$\xi=0$.

For the case of $R_{rr}=0$, substituting the expansions in Eq.~\rf{asyexp}
together with Eq.~\rf{case2brsol} into Eqs.~\rf{ssefe} and \rf{ssvfe}, we find
the following recurrence relations
\bea
&& \nu_1 = -\mu_1 , 
\nonumber \\
&& \nu_2 = -\frac{1}{3} \left( \mu_1^2 + 2\mu_2 \right) ,
\nonumber \\
&& \la_{2} = \frac{2\xi \la_0 }{3\ka} \left( \mu_1^2 - \mu_2 \right) ,
\nonumber \\
&& ... \, ,
\label{case2asyrec}
\eea
where 5 coefficients, $\mu_0, \, \mu_1, \, \mu_2, \, \la_0$ and $\la_1$, are
unfixed. In this case, the system of equations can be reduced to three
second-order ODEs, so there are 6 free coefficients in the solutions taking into
account the trivial coefficient $\nu_0$. The asymptotic expansion of $b_r$ can
also be obtained. It is
\bw
\bea
b_r^2 &=& \frac{e^{2\mu_0} \left( e^{\mu_0} -1 \right)}{\xi} + \frac{ e^{2 \mu_0} }{ 9\xi\mu_1 } \Big[ \frac{2\xi}{\ka} \left( \left( \ka + 4\xi \right) \mu_1^2 + (8\ka-4\xi) \mu_2 \right) \la_0^2 + 18\xi \mu_1 \la_0 \la_1 + 3 (2\xi-\ka) \la_1^2   
\nonumber \\
&& + 2 \left( 15 e^{2 \mu_0}-9 \right)\mu_1^2 - 12 e^{2 \mu_0} \mu_2 \Big] \frac{1}{r} + ... \, .
\label{case2brasyexp}
\eea
\ew  
For $b_r^2 \ge 0$ at $r \rightarrow \infty$, it implies a constraint 
\bea
\frac{e^{\mu_0}-1}{\xi} \ge 0 .
\eea
If $\mu_0=0$, then there is a constraint
\bea
&&  2 \left[ \left( 1 + \frac{4\xi}{\ka} \right) \mu_1^2 + \left(8-\frac{4\xi}{\ka}\right) \mu_2 \right] \la_0^2 + 18 \mu_1 \la_0 \la_1    
\nonumber \\
&& \hspace{1cm} + 3 \left(2-\frac{\ka}{\xi}\right) \la_1^2 + \frac{12}{\xi} \left( \mu_1^2 - \mu_2 \right) \ge 0 .
\label{brsqcon}
\eea

We point out that this family of solutions includes the analytical solution
given by Eqs.~\rf{asol1} and \rf{asol2}. It is recovered by taking $\mu_2 =
\mu_1^2$.

\section{Behavior of the variables near the event horizon}
\label{app3}

For the case of $b_r=0$, substituting the expansions in Eq.~\rf{hexp} into
Eqs.~\rf{ssefe} and \rf{ssvfe}, we find the following three equations relating
the 7 parameters $\big\{ r_h,\, N_{10}, \, N_{20}, \, M_{10}, \, M_{20}, \,
L_{10}, \, L_{20} \big\}$ if $N_{10} \ne 0$,
\bw
\bea
0 &=& -16 M_{10} N_{10}^3+\xi N_{20}^2 L_{10}^2 r_h^2-2 \xi N_{10} N_{20} L_{10} L_{20} r_h^2+2 \ka N_{10}^2 L_{20}^2 r_h^2, 
\nonumber \\
0 &=& 16 \ka M_{10} N_{10}^4-16 \ka N_{10}^4 r_h - 8\xi (\xi-2\ka)N_{10}^3 L_{10}^2 r_h+3 \ka \xi N_{10} N_{20}^2 L_{10}^2 r_h^2 + \xi^2 (\xi-2 \ka) N_{20}^2 L_{10}^4 r_h^2-6 \ka \xi N_{10}^2 N_{20} L_{10} L_{20} r_h^2
\nonumber \\
&& - 2\xi^2 (\xi-2\ka) N_{10} N_{20} L_{10}^3 L_{20} r_h^2 + 2\ka(2 \xi-\ka) N_{10}^3 L_{20}^2 r_h^2 + 2\ka\xi( \xi -2 \ka) N_{10}^2 L_{10}^2 L_{20}^2 r_h^2, 
\nonumber \\
0 &=& 64 \ka M_{10} M_{20} N_{10}^5 L_{10}+176 \ka M_{10}^2 N_{10}^4 N_{20} L_{10} + 16\xi(\xi-2 \ka)M_{10}^2 N_{10}^3 N_{20} L_{10}^3-128 \ka M_{10}^2 N_{10}^5 L_{20}-16 \ka M_{20} N_{10}^5 L_{10} r_h
\nonumber \\
&& - 176 \ka M_{10} N_{10}^4 N_{20} L_{10} r_h - 8\xi(\xi-2 \ka)M_{20} N_{10}^4 L_{10}^3 r_h - 56\xi (\xi-2\ka) M_{10} N_{10}^3 N_{20} L_{10}^3 r_h+128 \ka M_{10} N_{10}^5 L_{20} r_h
\nonumber \\
&& +12 \ka \xi M_{20} N_{10}^2 N_{20}^2 L_{10}^3 r_h^2+21 \ka \xi M_{10} N_{10} N_{20}^3 L_{10}^3 r_h^2 + 4\xi^2(\xi-2 \ka) M_{20} N_{10} N_{20}^2 L_{10}^5 r_h^2 + 5\xi^2(\xi-2\ka) M_{10} N_{20}^3 L_{10}^5 r_h^2
\nonumber \\
&& -24 \ka \xi M_{20} N_{10}^3 N_{20} L_{10}^2 L_{20} r_h^2-66 \ka \xi M_{10} N_{10}^2 N_{20}^2 L_{10}^2 L_{20} r_h^2 - 8\xi^2(\xi-2\ka) M_{20} N_{10}^2 N_{20} L_{10}^4 L_{20} r_h^2 
\nonumber \\
&& - 10\xi^2(\xi-2 \ka) M_{10} N_{10} N_{20}^2 L_{10}^4 L_{20} r_h^2 + 8\ka(2\xi-\ka) M_{20} N_{10}^4 L_{10} L_{20}^2 r_h^2 + 2\ka(40\xi-11\ka) M_{10} N_{10}^3 N_{20} L_{10} L_{20}^2 r_h^2
\nonumber \\
&& + 8\ka\xi(\xi-2\ka) M_{20} N_{10}^3 L_{10}^3 L_{20}^2 r_h^2 + 10\ka\xi(\xi-2 \ka) M_{10} N_{10}^2 N_{20} L_{10}^3 L_{20}^2 r_h^2 - 16\ka(2\xi-\ka) M_{10} N_{10}^4 L_{20}^3 r_h^2.
\label{case1receq1}
\eea
\ew
From Eq.~\rf{case1receq1}, three of the parameters can be solved in terms of the
other 4 parameters.  Other higher-order coefficients are also fixed by their
recurrence relations once the 4 free parameters are specified.  If $N_{10}=0$,
then substituting the expansions in Eq.~\rf{hexp} into Eqs.~\rf{ssefe} and
\rf{ssvfe} simply leads to $N_{2n}=M_{2n}=L_{2n}=0$ and $L_{10} = 0$. The
recurrence relations are much simplified with only three free parameters. Taking
the three free parameters as $N_{11}, \, L_{11}$ and $r_h$, the recurrence
relations for the first few coefficients are
\bw
\bea
M_{10} &=& r_h + \left( \frac{\ka}{2} - \frac{\xi}{4} \right) \frac{L_{11}^2 r_h^2}{N_{11}} ,
\nonumber \\
M_{11} &=& 1 - \left( \frac{\ka^2}{4}-\ka \xi+\frac{13 \xi^2}{16}-\frac{3 \xi^3}{16 \ka} \right) \frac{L_{11}^4 r_h^2}{N_{11}^2}
 + \left( \frac{3 \ka^2 \xi}{8}-\frac{9 \ka \xi^2}{16}+\frac{9 \xi^3}{32}-\frac{3 \xi^4}{64 \ka} \right) \frac{L_{11}^6 r_h^3}{N_{11}^3},
\nonumber \\
M_{12} &=& \left( \frac{\ka^2}{4}-\ka \xi+\frac{13 \xi^2}{16}-\frac{3 \xi^3}{16 \ka}\right) \frac{L_{11}^4 r_h}{N_{11}^2}
 + \left( \frac{\ka^3}{8}-\frac{19 \ka^2 \xi}{12}+\frac{105 \ka \xi^2}{32}-\frac{173 \xi^3}{64}+\frac{191 \xi^4}{192 \ka}-\frac{35 \xi^5}{256 \ka^2} \right) \frac{ L_{11}^6 r_h^2}{ N_{11}^3}
\nonumber \\
&& - \left( \frac{49 \ka^3 \xi}{96}-\frac{433 \ka^2 \xi^2 }{192 }+\frac{207 \ka \xi^3 }{64 }-\frac{809 \xi^4 }{384 }+\frac{997 \xi^5 }{1536 \ka }-\frac{79 \xi^6 }{1024 \ka^2 }\right) \frac{L_{11}^8 r_h^3}{N_{11}^4}
\nonumber \\
&& + \left( \frac{11 \ka^3 \xi^2 }{32 }-\frac{55 \ka^2 \xi^3 }{64 }+\frac{55 \ka \xi^4 }{64 }-\frac{55 \xi^5 }{128}+\frac{55 \xi^6}{512 \ka }-\frac{11 \xi^7 }{1024 \ka^2 } \right) \frac{L_{11}^{10} r_h^4}{N_{11}^5} ,
\nonumber \\
N_{12} &=& -\frac{N_{11}}{r_h} + \left( \frac{ \ka}{2}-\frac{ \xi}{4}\right) L_{11}^2 + \left( \frac{ \ka \xi}{4 }-\frac{ \xi^2}{4} + \frac{ \xi^3}{16 \ka}\right) \frac{L_{11}^4 r_h}{N_{11}} ,
\nonumber \\ 
L_{12} &=& -\frac{L_{11}}{r_h} + \left( \frac{ \xi}{4 } -\frac{ \xi^2}{8 \ka} \right) \frac{L_{11}^3}{N_{11}} + \left( \frac{ \ka\xi}{4 }-\frac{ \xi^2}{4 } + \frac{ \xi^3}{16 \ka} \right) \frac{L_{11}^5 r_h}{N_{11}^2}   .
\eea
\ew
Note that for $\xi \ne 2\ka$, the Schwarzschild metric is recovered only when
$L_{11}=0$ and therefore $b_t = 0$. For $\xi=2\ka$, the Schwarzschild metric can
exist together with a nonzero $b_t$.

For the case of $R_{rr}=0$, using Eq.~\rf{case2brsol} to eliminate $b_r$ and
then substituting the expansions in Eq.~\rf{hexp} into Eqs.~\rf{ssefe} and
\rf{ssvfe}, we find the following equation relating the 7 parameters $\big\{
r_h,\, N_{10}, \, N_{20}, \, M_{10}, \, M_{20}, \, L_{10}, \, L_{20}\big\}$ if
$N_{10} \ne 0$,
\bw
\bea
0 &=& 384 \ka M_{10}^2 N_{10}^3-36 \ka M_{10} M_{20} N_{10}^2 N_{20} r_h+12 \ka M_{10}^2 N_{10} N_{20}^2 r_h - 8\xi \left(2\xi-\ka\right) M_{10} N_{20}^2 L_{10}^2 r_h^2 - 48 \ka^2 M_{10} N_{10}^2 L_{20}^2 r_h^2
\nonumber \\
&& + 16\xi \left(2\xi-\ka\right) M_{10} N_{10} N_{20} L_{10} L_{20} r_h^2 - 3\ka\left(\xi-2\ka\right) M_{20} N_{10} N_{20} L_{20}^2 r_h^3 + \ka\left(\xi-2 \ka\right) M_{10} N_{20}^2 L_{20}^2 r_h^3  .
\label{case2receq1}
\eea
\ew
From Eq.~\rf{case2receq1}, one of the parameters can be solved in terms of the
other 6 parameters.  Other higher-order coefficients are then fixed by their
recurrence relations once the 6 free parameters are specified.  Similar to the
first case, if $N_{10}=0$ then recurrence relations give
$N_{2n}=M_{2n}=L_{2n}=0$, but in this case $L_{10}$ is no longer fixed to zero
so there are still 6 free parameters. Taking the 6 free parameters as $ N_{11},
\, M_{10}, \, M_{11}, \, L_{10}, \, L_{11}$ and $r_h$, the recurrence relations
for the first few coefficients are
\bw
\bea
M_{12} &=& \frac{ M_{10}-M_{11}r_h }{\ka-2 \xi} \Bigg[  \frac{10(13\ka-8\xi)}{27r_h^2} - \frac{ 10\ka M_{10} N_{11} }{ 3\xi L_{10}^2 r_h^2 }   
+ \frac{13(2\xi-\ka)}{27} \frac{M_{11}}{M_{10} r_h}
- \frac{40(\xi-2\ka)}{9} \frac{ L_{11} }{L_{10} r_h}
- \frac{10\ka(\xi-2\ka)}{9\xi} \frac{L_{11}^2 }{ L_{10}^2}
 \Bigg] , 
\nonumber \\
N_{12} &=& \frac{M_{11} N_{11}}{3 M_{10}} -\frac{4 N_{11}}{3 r_h}, 
\nonumber \\
L_{12} &=& \frac{M_{11} L_{11}}{3 M_{10}}-\frac{4 L_{11}}{3 r_h} + \frac{2\xi M_{11} L_{10}}{3\ka M_{10} r_h}-\frac{2\xi L_{10}}{3 \ka r_h^2} .
\eea
\ew
Note that the analytical solution in Eqs.~\rf{asol1} and \rf{asol2} corresponds
to taking 
\bea
&& r_h = 2M, \nonumber \\
&& N_{10}=0 , \nonumber \\
&& N_{11}=\frac{1}{2M}, \nonumber \\
&& M_{10} = 2M e^{2\mu_0}, \nonumber \\
&& M_{11} = e^{2\mu_0}, \nonumber \\
&& L_{10} = \la_0 + \frac{\la_1 }{2M} , \nonumber \\
&& L_{11} = - \frac{\la_1}{4M^2} .
\eea

\section{Curvature scalars at the event horizon}
\label{app4}

If $N_{10} \ne 0$, the metric expansion in Eq.~\rf{hexp} gives finite values for
the curvature scalars $R, \, R_{\al\be}R^{\al\be}$ and $R_{\abgd}R^{\abgd}$ when
$\de = r-r_h$ approaches zero. They are 
\bw
\bea
&& R |_{r=r_h} = \frac{2}{r_h^2} - \frac{2}{r_h M_{10} } + \frac{M_{20} N_{20}}{8 M_{10}^2 N_{10}}+\frac{N_{20}^2}{8 M_{10} N_{10}^2}-\frac{N_{11}}{2 M_{10} N_{10}},
\nonumber \\
&& R_{\al\be}R^{\al\be} |_{r=r_h} = \frac{2}{r_h^4} - \frac{2}{r_h^3 M_{10}} + \frac{3}{2 r_h^2 M_{10}^2} - \frac{M_{20} N_{20}}{8 r_h M_{10}^3 N_{10}} - \frac{N_{20}^2}{8 r_h M_{10}^2 N_{10}^2 } + \frac{M_{20}^2 N_{20}^2}{128 M_{10}^4 N_{10}^2}+\frac{M_{20} N_{20}^3}{64 M_{10}^3 N_{10}^3} + \frac{N_{20}^4}{128 M_{10}^2 N_{10}^4}
\nonumber \\
&& \hskip 1.8cm + \frac{N_{11}}{2 r_h M_{10}^2 N_{10} } - \frac{M_{20} N_{11} N_{20}}{16 M_{10}^3 N_{10}^2} - \frac{ N_{20}^2 N_{11} }{16 M_{10}^2 N_{10}^3}+\frac{N_{11}^2}{8 M_{10}^2 N_{10}^2},
\nonumber \\
&& R_{\abgd} R^{\abgd} |_{r=r_h} = \frac{4}{r_h^4} + \frac{2}{r_h^2 M_{10}^2} +\frac{M_{20} N_{20}^3}{32 M_{10}^3 N_{10}^3} + \frac{M_{20}^2 N_{20}^2}{64 M_{10}^4 N_{10}^2} +\frac{N_{20}^4}{64 M_{10}^2 N_{10}^4} - \frac{M_{20} N_{20} N_{11}}{8 M_{10}^3 N_{10}^2} - \frac{N_{20}^2 N_{11}}{8 M_{10}^2 N_{10}^3} + \frac{N_{11}^2}{4 M_{10}^2 N_{10}^2}.
\label{chcurvature}
\eea 
\ew
Note that $M_{10}$ is always nonzero in our numerical solutions. The relations
in Eqs.~\rf{case1receq1} and \rf{case2receq1} as well as the corresponding
recurrence relations for $N_{11}$ might be used to rewrite the results in
Eq.~\rf{chcurvature} for each family of solutions. While the rewriting does not
bring any simplification for the first family of solutions ($b_r=0$), for the
second family of solutions ($R_{rr}=0$), the results in Eq.~\rf{chcurvature}
remarkably simplify to
\bea
&& R |_{r=r_h} = \frac{2}{r_h^2},
\nonumber \\
&& R_{\al\be}R^{\al\be} |_{r=r_h} = \frac{2}{r_h^4} - \frac{2}{r_h^3 M_{10}} + \frac{3}{2 r_h^2 M_{10}^2},
\nonumber \\
&& R_{\abgd} R^{\abgd} |_{r=r_h} = \frac{4}{r_h^4} + \frac{6}{r_h^2 M_{10}^2}.
\eea 
The coordinate scalar $b^2 : = g_\mn b^\mu b^\nu$ is also finite at $r_h$. It is
\bea
b^2 |_{r=r_h} = -\frac{L_{10}^2}{N_{10}}
\eea
for the first family of solutions, and 
\bea
b^2 |_{r=r_h} = \frac{ 2M_{10}}{\xi r_h } - \frac{1}{\xi} + \frac{(\ka-2\xi) L_{10}^2}{3\ka N_{10} } + \frac{(\xi-2 \ka) r_h L_{20}^2 }{6 \xi N_{10} } 
\eea
for the second family of solutions.

If $N_{10} = 0$ and thus $N_{2n}=M_{2n}=L_{2n}=0$ in Eq.~\rf{hexp}, the finite
values for the curvature scalars $R, \, R_{\al\be}R^{\al\be}$ and
$R_{\abgd}R^{\abgd}$ are then
\bw
\bea
&& R |_{r=r_h} = \frac{2}{r_h^2} - \frac{4}{r_h M_{10} } + \frac{M_{11}}{2 M_{10}^2} - \frac{3 N_{12}}{2 M_{10} N_{11}},
\nonumber \\
&& R_{\al\be}R^{\al\be} |_{r=r_h} = \frac{2}{r_h^4} - \frac{4}{r_h^3 M_{10}} +\frac{4}{r_h^2 M_{10}^2} - \frac{M_{11}}{r_h M_{10}^3} + \frac{3 N_{12}}{r_h M_{10}^2 N_{11}}
+ \frac{M_{11}^2}{8 M_{10}^4}-\frac{3 M_{11} N_{12}}{4 M_{10}^3 N_{11}}+\frac{9 N_{12}^2}{8 M_{10}^2 N_{11}^2},
\nonumber \\
&& R_{\abgd} R^{\abgd} |_{r=r_h} = \frac{4}{r_h^4} +\frac{4}{r_h^2 M_{10}^2} + \frac{M_{11}^2}{4 M_{10}^4} - \frac{3 M_{11} N_{12}}{2 M_{10}^3 N_{11}} + \frac{9 N_{12}^2}{4 M_{10}^2 N_{11}^2},
\label{bhcurvature}
\eea
\ew
when $\de$ approaches zero. The recurrence relations for the first family of
solutions do not bring any simplification for the results in
Eq.~\rf{bhcurvature}. The recurrence relations for the second family of
solutions simplify the results to 
\bea
&& R |_{r=r_h} = \frac{2}{r_h^2} - \frac{2 }{r_h M_{10}},
\nonumber \\
&& R_{\al\be}R^{\al\be} |_{r=r_h} = \frac{2 }{r_h^4} - \frac{4}{r_h^3 M_{10}} + \frac{2}{r_h^2 M_{10}^2},
\nonumber \\
&& R_{\abgd} R^{\abgd} |_{r=r_h} = \frac{4}{r_h^4} + \frac{8}{r_h^2 M_{10}^2},
\eea
which recovers the results for the Schwarzschild metric by taking $M_{10}=r_h$.
With $N_{10}=0$, we have $b^2 |_{r=r_h} =0$ for the first family of solutions,
and 
\bea
b^2 |_{r=r_h} &=& \frac{M_{10}}{\xi r_h} - \frac{(4 \xi + 7 \ka) L_{10}^2}{9 \ka r_h N_{11}} + \frac{2(2\xi - \ka) M_{11} L_{10}^2 }{9 \ka M_{10} N_{11} } 
\nonumber \\
&&  + \frac{(\xi - 2\ka )r_h L_{11}^2 }{3 \xi N_{11}} -\frac{2 L_{10} L_{11}}{N_{11}} - \frac{1}{\xi}
\eea
for the second family of solutions.

\section{Asymptotic expansion of the metric in the harmonic coordinates}
\label{app5}

The PPN metric expansion is constructed in the harmonic coordinates. For static
spherical spacetime, it is \cite{Poisson:2014}
\bea
&& \bar g_{00} = -1 + \frac{2M}{\bar r} - 2 \be \left( \frac{M}{\bar r} \right)^2 + ... ,
\nonumber \\
&& \bar g_{ij} = \left( 1 + 2\ga \frac{M}{\bar r} \right) \de_{ij} + ..., 
\label{harmetric1}
\eea  
where $\bar r=\sqrt{\bar x^2 + \bar y^2 + \bar z^2}$ is the Euclidean length of
the position vector in the harmonic coordinates, and $\be$ and $\ga$ are the PPN
parameters that turn out to be $2+\nu_2/\mu_1^2$ and $1$ to match the expansion
in Eq.~\rf{asyexp} when $\mu_0=0$. In the following derivation, we show the
match by transforming the metric expansion in Eq.~\rf{asyexp} to the harmonic
coordinates. 
 
The transformation includes two steps: (i) from the coordinates $(t, \, r, \,
\th, \, \ph)$ to $(t, \, \bar r, \, \th, \, \ph)$, and (ii) then to the harmonic
coordinates $(t, \, \bar x, \, \bar y, \, \bar z)$. Denoting 
\bea
g_{\bar r \bar r} = e^{2\mu} \left( \frac{dr}{d\bar r} \right)^2,
\eea
the metric components in the harmonic coordinates are
\bea
&& \bar g_{00} = -e^{2\nu}, 
\nonumber \\
&& \bar g_{ij} = \frac{\prt \bar r}{\prt \bar x^i} \frac{\prt \bar r}{\prt \bar x^j} g_{\bar r \bar r} + \frac{\prt \th}{\prt \bar x^i} \frac{\prt \th}{\prt \bar x^j} r^2 + \frac{\prt \ph}{\prt \bar x^i} \frac{\prt \ph}{\prt \bar x^j} r^2 \sin^2\th ,
\label{harmetric2}
\eea 
where the Jacobian matrix for the transformation between $(\bar r, \, \th, \,
\ph)$ and $(\bar x, \, \bar y, \, \bar z)$ has the usual elements

\bea
\begin{pmatrix}
\frac{\prt \bar r}{\prt \bar x} & \frac{\prt \bar r}{\prt \bar y} & \frac{\prt \bar r}{\prt \bar z} \\
\frac{\prt \th}{\prt \bar x} & \frac{\prt \th}{\prt \bar y} & \frac{\prt \th}{\prt \bar z} \\
\frac{\prt \ph}{\prt \bar x} & \frac{\prt \ph}{\prt \bar y} & \frac{\prt \ph}{\prt \bar z}
\end{pmatrix}
= \begin{pmatrix}
 \sin\th\cos\ph &  \sin\th\sin\ph & \cos\th\\
 \frac{1}{\bar r}\cos\th\cos\ph  & \frac{1}{\bar r}\cos\th\sin\ph & -\frac{1}{\bar r}\sin\th \\
 -\frac{1}{\bar r}\frac{\sin\ph}{\sin\th} & \frac{1}{\bar r}\frac{\cos\ph}{\sin\th} & 0 
\end{pmatrix} .
\eea

Calculating the inverse metric $\bar g^{\al\be}$ and the Christoffel symbols
$\bar \Ga^{\la}_{\pt \mu \al\be}$ in the harmonic coordinates, we find that the
harmonic condition 
\bea
\bar g^{\al\be} \, \bar \Ga^{\la}_{\pt\la \al\be} = 0
\eea
simplifies to an ODE for $\bar r$ as a function of $r$, 
\bea
\bar r'' + \left(\nu'-\mu'+\frac{2}{r}\right) \bar r' - \frac{2 e^{2\mu} }{r^2} \bar r = 0 ,
\label{harmoniceq}
\eea 
where the prime denotes the derivative with respect to $r$. 
With $\mu$ and $\nu$ expressed as the expansions in Eq.~\rf{asyexp},
Eq.~\rf{harmoniceq} has the following asymptotic solutions
\bea
\bar r = r^s \left( 1 + \frac{c_1}{r} + \frac{c_2}{r^2} + ... \right) ,
\label{transol}
\eea
where the constant $s$ and the coefficients $c_1, \, c_2, \, ...$, can be
determined by substituting the solution back into Eq.~\rf{harmoniceq}. We find
the indicial equation for $s$ to be
\bea
s^2 + s - 2 e^{2\mu_0} = 0,
\label{indicialeq}
\eea
and the coefficients to be
\bea
&& c_1 = \frac{s(\mu_1-\nu_1)-4e^{2\mu_0} \mu_1}{2s},
\nonumber \\
&& c_2 = \frac{ e^{2 \mu_0} \left[ 4 e^{2 \mu_0} \mu_1^2 + (1-4s) \mu_1^2 + (2s-1) \mu_1 \nu_1 - 2 s\mu_2 \right] }{ s (2 s-1)} 
\nonumber \\
&& \hskip 0.7cm  + \frac{ (s-1)(\mu_1 - \nu_1)^2 + 4 s (\mu_2-\nu_2)}{4 (2 s-1)},
\nonumber \\
&& ... \, .
\label{transoliter}
\eea
Note that the two roots of Eq.~\rf{indicialeq} give two independent solutions whose linear combinations produce general solutions to Eq.~\rf{harmoniceq}. 
But for our purpose, it is sensible to have $s=1$ when $\mu_0=0$, suggesting to take the special solution corresponding to the root 
\bea
s= \frac{-1+\sqrt{1+8e^{2\mu_0}}}{2} .
\eea

To eliminate $r$ in Eq.~\rf{harmetric2} so that it can be compared with
Eq.~\rf{harmetric1}, the inverse of Eq.~\rf{transol} is required. We find
\bea
r = u \left( 1 - \frac{c_1}{s}\frac{1}{u} - \frac{c_2+\frac{1-s}{2s}c_1^2}{s} \frac{1}{u^2} + ... \right),
\label{transolinverse}
\eea 
where $u = \bar r^{1/s}$. Focusing on the transformation when $\mu_0=0$, we have
$s=1$ and Eq.~\rf{transolinverse} is simplified to
\bea
r = \bar r \left( 1 + \frac{\mu_1}{\bar r} + \frac{ \mu_2 + \nu_2 }{{\bar r}^2} + ... \right) ,
\label{transolinverse2}
\eea
where the relation $\nu_1=-\mu_1$, which is valid for both families of
solutions, has also been used.  Using Eq.~\rf{transolinverse2} to eliminate $r$
in Eq.~\rf{harmetric2} then leads to 
\bea
&& \bar g_{00} = - 1 + \frac{2\mu_1}{\bar r} - \frac{4\mu_1^2+2\nu_2}{\bar r^2} + ... , 
\nonumber \\
&& \bar g_{ij} = \left( 1 + \frac{2\mu_1}{\bar r} \right) \de_{ij} + ... \, .
\eea
With $\mu_1=M$, we see 
\bea
\be = 2+\frac{\nu_2}{\mu_1^2}, \quad \ga = 1.
\eea
Using the recurrence relations for $\nu_2$ in Eqs.~\rf{case1asyrec} and
\rf{case2asyrec}, $\be$ can be related to the free coefficients in each family
of solutions.

\bibliography{refs}

\begin{thebibliography}{41}%
\makeatletter
\providecommand \@ifxundefined [1]{%
 \@ifx{#1\undefined}
}%
\providecommand \@ifnum [1]{%
 \ifnum #1\expandafter \@firstoftwo
 \else \expandafter \@secondoftwo
 \fi
}%
\providecommand \@ifx [1]{%
 \ifx #1\expandafter \@firstoftwo
 \else \expandafter \@secondoftwo
 \fi
}%
\providecommand \natexlab [1]{#1}%
\providecommand \enquote  [1]{``#1''}%
\providecommand \bibnamefont  [1]{#1}%
\providecommand \bibfnamefont [1]{#1}%
\providecommand \citenamefont [1]{#1}%
\providecommand \href@noop [0]{\@secondoftwo}%
\providecommand \href [0]{\begingroup \@sanitize@url \@href}%
\providecommand \@href[1]{\@@startlink{#1}\@@href}%
\providecommand \@@href[1]{\endgroup#1\@@endlink}%
\providecommand \@sanitize@url [0]{\catcode `\\12\catcode `\$12\catcode
  `\&12\catcode `\#12\catcode `\^12\catcode `\_12\catcode `\%12\relax}%
\providecommand \@@startlink[1]{}%
\providecommand \@@endlink[0]{}%
\providecommand \url  [0]{\begingroup\@sanitize@url \@url }%
\providecommand \@url [1]{\endgroup\@href {#1}{\urlprefix }}%
\providecommand \urlprefix  [0]{URL }%
\providecommand \Eprint [0]{\href }%
\providecommand \doibase [0]{http://dx.doi.org/}%
\providecommand \selectlanguage [0]{\@gobble}%
\providecommand \bibinfo  [0]{\@secondoftwo}%
\providecommand \bibfield  [0]{\@secondoftwo}%
\providecommand \translation [1]{[#1]}%
\providecommand \BibitemOpen [0]{}%
\providecommand \bibitemStop [0]{}%
\providecommand \bibitemNoStop [0]{.\EOS\space}%
\providecommand \EOS [0]{\spacefactor3000\relax}%
\providecommand \BibitemShut  [1]{\csname bibitem#1\endcsname}%
\let\auto@bib@innerbib\@empty
\bibitem [{\citenamefont {Will}\ and\ \citenamefont
  {Nordtvedt}(1972)}]{Will:1972zz}%
  \BibitemOpen
  \bibfield  {author} {\bibinfo {author} {\bibfnamefont {C.~M.}\ \bibnamefont
  {Will}}\ and\ \bibinfo {author} {\bibfnamefont {K.}~\bibnamefont {Nordtvedt},
  \bibfnamefont {Jr.}},\ }\href {\doibase 10.1086/151754} {\bibfield  {journal}
  {\bibinfo  {journal} {Astrophys. J.}\ }\textbf {\bibinfo {volume} {177}},\
  \bibinfo {pages} {757} (\bibinfo {year} {1972})}\BibitemShut {NoStop}%
\bibitem [{\citenamefont {Kosteleck\'y}\ and\ \citenamefont
  {Samuel}(1989{\natexlab{a}})}]{Kostelecky:1988zi}%
  \BibitemOpen
  \bibfield  {author} {\bibinfo {author} {\bibfnamefont {V.~A.}\ \bibnamefont
  {Kosteleck\'y}}\ and\ \bibinfo {author} {\bibfnamefont {S.}~\bibnamefont
  {Samuel}},\ }\href {\doibase 10.1103/PhysRevD.39.683} {\bibfield  {journal}
  {\bibinfo  {journal} {Phys. Rev. D}\ }\textbf {\bibinfo {volume} {39}},\
  \bibinfo {pages} {683} (\bibinfo {year} {1989}{\natexlab{a}})}\BibitemShut
  {NoStop}%
\bibitem [{\citenamefont {Kosteleck\'y}\ and\ \citenamefont
  {Samuel}(1989{\natexlab{b}})}]{Kostelecky:1989jp}%
  \BibitemOpen
  \bibfield  {author} {\bibinfo {author} {\bibfnamefont {V.~A.}\ \bibnamefont
  {Kosteleck\'y}}\ and\ \bibinfo {author} {\bibfnamefont {S.}~\bibnamefont
  {Samuel}},\ }\href {\doibase 10.1103/PhysRevLett.63.224} {\bibfield
  {journal} {\bibinfo  {journal} {Phys. Rev. Lett.}\ }\textbf {\bibinfo
  {volume} {63}},\ \bibinfo {pages} {224} (\bibinfo {year}
  {1989}{\natexlab{b}})}\BibitemShut {NoStop}%
\bibitem [{\citenamefont {Colladay}\ and\ \citenamefont
  {Kosteleck\'y}(1997)}]{Colladay:1996iz}%
  \BibitemOpen
  \bibfield  {author} {\bibinfo {author} {\bibfnamefont {D.}~\bibnamefont
  {Colladay}}\ and\ \bibinfo {author} {\bibfnamefont {V.~A.}\ \bibnamefont
  {Kosteleck\'y}},\ }\href {\doibase 10.1103/PhysRevD.55.6760} {\bibfield
  {journal} {\bibinfo  {journal} {Phys. Rev. D}\ }\textbf {\bibinfo {volume}
  {55}},\ \bibinfo {pages} {6760} (\bibinfo {year} {1997})},\ \Eprint
  {http://arxiv.org/abs/hep-ph/9703464} {arXiv:hep-ph/9703464} \BibitemShut
  {NoStop}%
\bibitem [{\citenamefont {Colladay}\ and\ \citenamefont
  {Kosteleck\'y}(1998)}]{Colladay:1998fq}%
  \BibitemOpen
  \bibfield  {author} {\bibinfo {author} {\bibfnamefont {D.}~\bibnamefont
  {Colladay}}\ and\ \bibinfo {author} {\bibfnamefont {V.~A.}\ \bibnamefont
  {Kosteleck\'y}},\ }\href {\doibase 10.1103/PhysRevD.58.116002} {\bibfield
  {journal} {\bibinfo  {journal} {Phys. Rev. D}\ }\textbf {\bibinfo {volume}
  {58}},\ \bibinfo {pages} {116002} (\bibinfo {year} {1998})},\ \Eprint
  {http://arxiv.org/abs/hep-ph/9809521} {arXiv:hep-ph/9809521} \BibitemShut
  {NoStop}%
\bibitem [{\citenamefont {Kosteleck\'y}(2004)}]{Kostelecky:2003fs}%
  \BibitemOpen
  \bibfield  {author} {\bibinfo {author} {\bibfnamefont {V.~A.}\ \bibnamefont
  {Kosteleck\'y}},\ }\href {\doibase 10.1103/PhysRevD.69.105009} {\bibfield
  {journal} {\bibinfo  {journal} {Phys. Rev. D}\ }\textbf {\bibinfo {volume}
  {69}},\ \bibinfo {pages} {105009} (\bibinfo {year} {2004})},\ \Eprint
  {http://arxiv.org/abs/hep-th/0312310} {arXiv:hep-th/0312310} \BibitemShut
  {NoStop}%
\bibitem [{\citenamefont {Kosteleck\'y}\ and\ \citenamefont
  {Mewes}(2009)}]{Kostelecky:2009zp}%
  \BibitemOpen
  \bibfield  {author} {\bibinfo {author} {\bibfnamefont {V.~A.}\ \bibnamefont
  {Kosteleck\'y}}\ and\ \bibinfo {author} {\bibfnamefont {M.}~\bibnamefont
  {Mewes}},\ }\href {\doibase 10.1103/PhysRevD.80.015020} {\bibfield  {journal}
  {\bibinfo  {journal} {Phys. Rev. D}\ }\textbf {\bibinfo {volume} {80}},\
  \bibinfo {pages} {015020} (\bibinfo {year} {2009})},\ \Eprint
  {http://arxiv.org/abs/0905.0031} {arXiv:0905.0031 [hep-ph]} \BibitemShut
  {NoStop}%
\bibitem [{\citenamefont {Kosteleck\'y}\ and\ \citenamefont
  {Mewes}(2012)}]{Kostelecky:2011gq}%
  \BibitemOpen
  \bibfield  {author} {\bibinfo {author} {\bibfnamefont {V.~A.}\ \bibnamefont
  {Kosteleck\'y}}\ and\ \bibinfo {author} {\bibfnamefont {M.}~\bibnamefont
  {Mewes}},\ }\href {\doibase 10.1103/PhysRevD.85.096005} {\bibfield  {journal}
  {\bibinfo  {journal} {Phys. Rev. D}\ }\textbf {\bibinfo {volume} {85}},\
  \bibinfo {pages} {096005} (\bibinfo {year} {2012})},\ \Eprint
  {http://arxiv.org/abs/1112.6395} {arXiv:1112.6395 [hep-ph]} \BibitemShut
  {NoStop}%
\bibitem [{\citenamefont {Kosteleck\'y}\ and\ \citenamefont
  {Mewes}(2013)}]{Kostelecky:2013rta}%
  \BibitemOpen
  \bibfield  {author} {\bibinfo {author} {\bibfnamefont {V.~A.}\ \bibnamefont
  {Kosteleck\'y}}\ and\ \bibinfo {author} {\bibfnamefont {M.}~\bibnamefont
  {Mewes}},\ }\href {\doibase 10.1103/PhysRevD.88.096006} {\bibfield  {journal}
  {\bibinfo  {journal} {Phys. Rev. D}\ }\textbf {\bibinfo {volume} {88}},\
  \bibinfo {pages} {096006} (\bibinfo {year} {2013})},\ \Eprint
  {http://arxiv.org/abs/1308.4973} {arXiv:1308.4973 [hep-ph]} \BibitemShut
  {NoStop}%
\bibitem [{\citenamefont {Kosteleck\'y}\ and\ \citenamefont
  {Li}(2019)}]{Kostelecky:2018yfa}%
  \BibitemOpen
  \bibfield  {author} {\bibinfo {author} {\bibfnamefont {V.~A.}\ \bibnamefont
  {Kosteleck\'y}}\ and\ \bibinfo {author} {\bibfnamefont {Z.}~\bibnamefont
  {Li}},\ }\href {\doibase 10.1103/PhysRevD.99.056016} {\bibfield  {journal}
  {\bibinfo  {journal} {Phys. Rev. D}\ }\textbf {\bibinfo {volume} {99}},\
  \bibinfo {pages} {056016} (\bibinfo {year} {2019})},\ \Eprint
  {http://arxiv.org/abs/1812.11672} {arXiv:1812.11672 [hep-ph]} \BibitemShut
  {NoStop}%
\bibitem [{\citenamefont {Kosteleck\'y}\ and\ \citenamefont
  {Russell}(2011)}]{Kostelecky:2008ts}%
  \BibitemOpen
  \bibfield  {author} {\bibinfo {author} {\bibfnamefont {V.~A.}\ \bibnamefont
  {Kosteleck\'y}}\ and\ \bibinfo {author} {\bibfnamefont {N.}~\bibnamefont
  {Russell}},\ }\href {\doibase 10.1103/RevModPhys.83.11} {\bibfield  {journal}
  {\bibinfo  {journal} {Rev. Mod. Phys.}\ }\textbf {\bibinfo {volume} {83}},\
  \bibinfo {pages} {11} (\bibinfo {year} {2011})},\ \Eprint
  {http://arxiv.org/abs/0801.0287} {arXiv:0801.0287 [hep-ph]} \BibitemShut
  {NoStop}%
\bibitem [{\citenamefont {Tasson}(2016)}]{Tasson:2016xib}%
  \BibitemOpen
  \bibfield  {author} {\bibinfo {author} {\bibfnamefont {J.~D.}\ \bibnamefont
  {Tasson}},\ }\href@noop {} {\bibfield  {journal} {\bibinfo  {journal}
  {Symmetry}\ }\textbf {\bibinfo {volume} {8}},\ \bibinfo {pages} {111}
  (\bibinfo {year} {2016})},\ \Eprint {http://arxiv.org/abs/1610.05357}
  {arXiv:1610.05357 [gr-qc]} \BibitemShut {NoStop}%
\bibitem [{\citenamefont {Bailey}\ and\ \citenamefont
  {Kosteleck\'y}(2006)}]{Bailey:2006fd}%
  \BibitemOpen
  \bibfield  {author} {\bibinfo {author} {\bibfnamefont {Q.~G.}\ \bibnamefont
  {Bailey}}\ and\ \bibinfo {author} {\bibfnamefont {V.~A.}\ \bibnamefont
  {Kosteleck\'y}},\ }\href {\doibase 10.1103/PhysRevD.74.045001} {\bibfield
  {journal} {\bibinfo  {journal} {Phys. Rev. D}\ }\textbf {\bibinfo {volume}
  {74}},\ \bibinfo {pages} {045001} (\bibinfo {year} {2006})},\ \Eprint
  {http://arxiv.org/abs/gr-qc/0603030} {arXiv:gr-qc/0603030} \BibitemShut
  {NoStop}%
\bibitem [{\citenamefont {Bluhm}\ and\ \citenamefont
  {Kosteleck\'y}(2005)}]{Bluhm:2004ep}%
  \BibitemOpen
  \bibfield  {author} {\bibinfo {author} {\bibfnamefont {R.}~\bibnamefont
  {Bluhm}}\ and\ \bibinfo {author} {\bibfnamefont {V.~A.}\ \bibnamefont
  {Kosteleck\'y}},\ }\href {\doibase 10.1103/PhysRevD.71.065008} {\bibfield
  {journal} {\bibinfo  {journal} {Phys. Rev. D}\ }\textbf {\bibinfo {volume}
  {71}},\ \bibinfo {pages} {065008} (\bibinfo {year} {2005})},\ \Eprint
  {http://arxiv.org/abs/hep-th/0412320} {arXiv:hep-th/0412320} \BibitemShut
  {NoStop}%
\bibitem [{\citenamefont {Bluhm}(2006)}]{Bluhm:2005uj}%
  \BibitemOpen
  \bibfield  {author} {\bibinfo {author} {\bibfnamefont {R.}~\bibnamefont
  {Bluhm}},\ }\href {\doibase 10.1007/3-540-34523-X_8} {\bibfield  {journal}
  {\bibinfo  {journal} {Lect. Notes Phys.}\ }\textbf {\bibinfo {volume}
  {702}},\ \bibinfo {pages} {191} (\bibinfo {year} {2006})},\ \Eprint
  {http://arxiv.org/abs/hep-ph/0506054} {arXiv:hep-ph/0506054} \BibitemShut
  {NoStop}%
\bibitem [{\citenamefont {Bluhm}\ \emph
  {et~al.}(2008{\natexlab{a}})\citenamefont {Bluhm}, \citenamefont {Fung},\
  and\ \citenamefont {Kosteleck\'y}}]{Bluhm:2007bd}%
  \BibitemOpen
  \bibfield  {author} {\bibinfo {author} {\bibfnamefont {R.}~\bibnamefont
  {Bluhm}}, \bibinfo {author} {\bibfnamefont {S.-H.}\ \bibnamefont {Fung}}, \
  and\ \bibinfo {author} {\bibfnamefont {V.~A.}\ \bibnamefont {Kosteleck\'y}},\
  }\href {\doibase 10.1103/PhysRevD.77.065020} {\bibfield  {journal} {\bibinfo
  {journal} {Phys. Rev. D}\ }\textbf {\bibinfo {volume} {77}},\ \bibinfo
  {pages} {065020} (\bibinfo {year} {2008}{\natexlab{a}})},\ \Eprint
  {http://arxiv.org/abs/0712.4119} {arXiv:0712.4119 [hep-th]} \BibitemShut
  {NoStop}%
\bibitem [{\citenamefont {Bluhm}\ \emph
  {et~al.}(2008{\natexlab{b}})\citenamefont {Bluhm}, \citenamefont {Gagne},
  \citenamefont {Potting},\ and\ \citenamefont {Vrublevskis}}]{Bluhm:2008yt}%
  \BibitemOpen
  \bibfield  {author} {\bibinfo {author} {\bibfnamefont {R.}~\bibnamefont
  {Bluhm}}, \bibinfo {author} {\bibfnamefont {N.~L.}\ \bibnamefont {Gagne}},
  \bibinfo {author} {\bibfnamefont {R.}~\bibnamefont {Potting}}, \ and\
  \bibinfo {author} {\bibfnamefont {A.}~\bibnamefont {Vrublevskis}},\ }\href
  {\doibase 10.1103/PhysRevD.79.029902} {\bibfield  {journal} {\bibinfo
  {journal} {Phys. Rev. D}\ }\textbf {\bibinfo {volume} {77}},\ \bibinfo
  {pages} {125007} (\bibinfo {year} {2008}{\natexlab{b}})},\ \bibinfo {note}
  {[Erratum: Phys.Rev.D 79, 029902(E)]},\ \Eprint
  {http://arxiv.org/abs/0802.4071} {arXiv:0802.4071 [hep-th]} \BibitemShut
  {NoStop}%
\bibitem [{\citenamefont {Liang}\ \emph {et~al.}(2022)\citenamefont {Liang},
  \citenamefont {Xu}, \citenamefont {Lu},\ and\ \citenamefont
  {Shao}}]{Liang:2022hxd}%
  \BibitemOpen
  \bibfield  {author} {\bibinfo {author} {\bibfnamefont {D.}~\bibnamefont
  {Liang}}, \bibinfo {author} {\bibfnamefont {R.}~\bibnamefont {Xu}}, \bibinfo
  {author} {\bibfnamefont {X.}~\bibnamefont {Lu}}, \ and\ \bibinfo {author}
  {\bibfnamefont {L.}~\bibnamefont {Shao}},\ }\href@noop {} {\  (\bibinfo
  {year} {2022})},\ \Eprint {http://arxiv.org/abs/2207.14423} {arXiv:2207.14423
  [gr-qc]} \BibitemShut {NoStop}%
\bibitem [{\citenamefont {Hellings}\ and\ \citenamefont
  {Nordtvedt}(1973)}]{Hellings:1973zz}%
  \BibitemOpen
  \bibfield  {author} {\bibinfo {author} {\bibfnamefont {R.~W.}\ \bibnamefont
  {Hellings}}\ and\ \bibinfo {author} {\bibfnamefont {K.}~\bibnamefont
  {Nordtvedt}},\ }\href {\doibase 10.1103/PhysRevD.7.3593} {\bibfield
  {journal} {\bibinfo  {journal} {Phys. Rev. D}\ }\textbf {\bibinfo {volume}
  {7}},\ \bibinfo {pages} {3593} (\bibinfo {year} {1973})}\BibitemShut
  {NoStop}%
\bibitem [{\citenamefont {Abbott}\ \emph {et~al.}(2019)\citenamefont {Abbott}
  \emph {et~al.}}]{LIGOScientific:2018mvr}%
  \BibitemOpen
  \bibfield  {author} {\bibinfo {author} {\bibfnamefont {B.~P.}\ \bibnamefont
  {Abbott}} \emph {et~al.} (\bibinfo {collaboration} {LIGO Scientific,
  Virgo}),\ }\href {\doibase 10.1103/PhysRevX.9.031040} {\bibfield  {journal}
  {\bibinfo  {journal} {Phys. Rev. X}\ }\textbf {\bibinfo {volume} {9}},\
  \bibinfo {pages} {031040} (\bibinfo {year} {2019})},\ \Eprint
  {http://arxiv.org/abs/1811.12907} {arXiv:1811.12907 [astro-ph.HE]}
  \BibitemShut {NoStop}%
\bibitem [{\citenamefont {Abbott}\ \emph
  {et~al.}(2021{\natexlab{a}})\citenamefont {Abbott} \emph
  {et~al.}}]{LIGOScientific:2020ibl}%
  \BibitemOpen
  \bibfield  {author} {\bibinfo {author} {\bibfnamefont {R.}~\bibnamefont
  {Abbott}} \emph {et~al.} (\bibinfo {collaboration} {LIGO Scientific,
  Virgo}),\ }\href {\doibase 10.1103/PhysRevX.11.021053} {\bibfield  {journal}
  {\bibinfo  {journal} {Phys. Rev. X}\ }\textbf {\bibinfo {volume} {11}},\
  \bibinfo {pages} {021053} (\bibinfo {year} {2021}{\natexlab{a}})},\ \Eprint
  {http://arxiv.org/abs/2010.14527} {arXiv:2010.14527 [gr-qc]} \BibitemShut
  {NoStop}%
\bibitem [{\citenamefont {Abbott}\ \emph
  {et~al.}(2021{\natexlab{b}})\citenamefont {Abbott} \emph
  {et~al.}}]{LIGOScientific:2021djp}%
  \BibitemOpen
  \bibfield  {author} {\bibinfo {author} {\bibfnamefont {R.}~\bibnamefont
  {Abbott}} \emph {et~al.} (\bibinfo {collaboration} {LIGO Scientific, VIRGO,
  KAGRA}),\ }\href@noop {} {\  (\bibinfo {year} {2021}{\natexlab{b}})},\
  \Eprint {http://arxiv.org/abs/2111.03606} {arXiv:2111.03606 [gr-qc]}
  \BibitemShut {NoStop}%
\bibitem [{\citenamefont {Akiyama}\ \emph {et~al.}(2019)\citenamefont {Akiyama}
  \emph {et~al.}}]{EventHorizonTelescope:2019dse}%
  \BibitemOpen
  \bibfield  {author} {\bibinfo {author} {\bibfnamefont {K.}~\bibnamefont
  {Akiyama}} \emph {et~al.} (\bibinfo {collaboration} {Event Horizon
  Telescope}),\ }\href {\doibase 10.3847/2041-8213/ab0ec7} {\bibfield
  {journal} {\bibinfo  {journal} {Astrophys. J. Lett.}\ }\textbf {\bibinfo
  {volume} {875}},\ \bibinfo {pages} {L1} (\bibinfo {year} {2019})},\ \Eprint
  {http://arxiv.org/abs/1906.11238} {arXiv:1906.11238 [astro-ph.GA]}
  \BibitemShut {NoStop}%
\bibitem [{\citenamefont {Akiyama}\ \emph
  {et~al.}(2022{\natexlab{a}})\citenamefont {Akiyama} \emph
  {et~al.}}]{EventHorizonTelescope:2022xnr}%
  \BibitemOpen
  \bibfield  {author} {\bibinfo {author} {\bibfnamefont {K.}~\bibnamefont
  {Akiyama}} \emph {et~al.} (\bibinfo {collaboration} {Event Horizon
  Telescope}),\ }\href {\doibase 10.3847/2041-8213/ac6674} {\bibfield
  {journal} {\bibinfo  {journal} {Astrophys. J. Lett.}\ }\textbf {\bibinfo
  {volume} {930}},\ \bibinfo {pages} {L12} (\bibinfo {year}
  {2022}{\natexlab{a}})}\BibitemShut {NoStop}%
\bibitem [{\citenamefont {Akiyama}\ \emph
  {et~al.}(2022{\natexlab{b}})\citenamefont {Akiyama} \emph
  {et~al.}}]{EventHorizonTelescope:2022xqj}%
  \BibitemOpen
  \bibfield  {author} {\bibinfo {author} {\bibfnamefont {K.}~\bibnamefont
  {Akiyama}} \emph {et~al.} (\bibinfo {collaboration} {Event Horizon
  Telescope}),\ }\href {\doibase 10.3847/2041-8213/ac6756} {\bibfield
  {journal} {\bibinfo  {journal} {Astrophys. J. Lett.}\ }\textbf {\bibinfo
  {volume} {930}},\ \bibinfo {pages} {L17} (\bibinfo {year}
  {2022}{\natexlab{b}})}\BibitemShut {NoStop}%
\bibitem [{\citenamefont {Casana}\ \emph {et~al.}(2018)\citenamefont {Casana},
  \citenamefont {Cavalcante}, \citenamefont {Poulis},\ and\ \citenamefont
  {Santos}}]{Casana:2017jkc}%
  \BibitemOpen
  \bibfield  {author} {\bibinfo {author} {\bibfnamefont {R.}~\bibnamefont
  {Casana}}, \bibinfo {author} {\bibfnamefont {A.}~\bibnamefont {Cavalcante}},
  \bibinfo {author} {\bibfnamefont {F.~P.}\ \bibnamefont {Poulis}}, \ and\
  \bibinfo {author} {\bibfnamefont {E.~B.}\ \bibnamefont {Santos}},\ }\href
  {\doibase 10.1103/PhysRevD.97.104001} {\bibfield  {journal} {\bibinfo
  {journal} {Phys. Rev. D}\ }\textbf {\bibinfo {volume} {97}},\ \bibinfo
  {pages} {104001} (\bibinfo {year} {2018})},\ \Eprint
  {http://arxiv.org/abs/1711.02273} {arXiv:1711.02273 [gr-qc]} \BibitemShut
  {NoStop}%
\bibitem [{\citenamefont {Cardoso}\ and\ \citenamefont
  {Pani}(2017)}]{Cardoso:2017cqb}%
  \BibitemOpen
  \bibfield  {author} {\bibinfo {author} {\bibfnamefont {V.}~\bibnamefont
  {Cardoso}}\ and\ \bibinfo {author} {\bibfnamefont {P.}~\bibnamefont {Pani}},\
  }\href {\doibase 10.1038/s41550-017-0225-y} {\bibfield  {journal} {\bibinfo
  {journal} {Nature Astron.}\ }\textbf {\bibinfo {volume} {1}},\ \bibinfo
  {pages} {586} (\bibinfo {year} {2017})},\ \Eprint
  {http://arxiv.org/abs/1709.01525} {arXiv:1709.01525 [gr-qc]} \BibitemShut
  {NoStop}%
\bibitem [{\citenamefont {Fan}(2018)}]{Fan:2017bka}%
  \BibitemOpen
  \bibfield  {author} {\bibinfo {author} {\bibfnamefont {Z.-Y.}\ \bibnamefont
  {Fan}},\ }\href {\doibase 10.1140/epjc/s10052-018-5540-7} {\bibfield
  {journal} {\bibinfo  {journal} {Eur. Phys. J. C}\ }\textbf {\bibinfo {volume}
  {78}},\ \bibinfo {pages} {65} (\bibinfo {year} {2018})},\ \Eprint
  {http://arxiv.org/abs/1709.04392} {arXiv:1709.04392 [hep-th]} \BibitemShut
  {NoStop}%
\bibitem [{\citenamefont {Arnowitt}\ \emph {et~al.}(2008)\citenamefont
  {Arnowitt}, \citenamefont {Deser},\ and\ \citenamefont
  {Misner}}]{Arnowitt:1962hi}%
  \BibitemOpen
  \bibfield  {author} {\bibinfo {author} {\bibfnamefont {R.~L.}\ \bibnamefont
  {Arnowitt}}, \bibinfo {author} {\bibfnamefont {S.}~\bibnamefont {Deser}}, \
  and\ \bibinfo {author} {\bibfnamefont {C.~W.}\ \bibnamefont {Misner}},\
  }\href {\doibase 10.1007/s10714-008-0661-1} {\bibfield  {journal} {\bibinfo
  {journal} {Gen. Rel. Grav.}\ }\textbf {\bibinfo {volume} {40}},\ \bibinfo
  {pages} {1997} (\bibinfo {year} {2008})},\ \Eprint
  {http://arxiv.org/abs/gr-qc/0405109} {arXiv:gr-qc/0405109} \BibitemShut
  {NoStop}%
\bibitem [{\citenamefont {Poisson}(2009)}]{Poisson:2009pwt}%
  \BibitemOpen
  \bibfield  {author} {\bibinfo {author} {\bibfnamefont {E.}~\bibnamefont
  {Poisson}},\ }\href {\doibase 10.1017/CBO9780511606601} {\emph {\bibinfo
  {title} {{A Relativist's Toolkit: The Mathematics of Black-Hole
  Mechanics}}}}\ (\bibinfo  {publisher} {Cambridge University Press},\ \bibinfo
  {year} {2009})\BibitemShut {NoStop}%
\bibitem [{\citenamefont {Westerweck}\ \emph {et~al.}(2018)\citenamefont
  {Westerweck}, \citenamefont {Nielsen}, \citenamefont {Fischer-Birnholtz},
  \citenamefont {Cabero}, \citenamefont {Capano}, \citenamefont {Dent},
  \citenamefont {Krishnan}, \citenamefont {Meadors},\ and\ \citenamefont
  {Nitz}}]{Westerweck:2017hus}%
  \BibitemOpen
  \bibfield  {author} {\bibinfo {author} {\bibfnamefont {J.}~\bibnamefont
  {Westerweck}}, \bibinfo {author} {\bibfnamefont {A.~B.}\ \bibnamefont
  {Nielsen}}, \bibinfo {author} {\bibfnamefont {O.}~\bibnamefont
  {Fischer-Birnholtz}}, \bibinfo {author} {\bibfnamefont {M.}~\bibnamefont
  {Cabero}}, \bibinfo {author} {\bibfnamefont {C.}~\bibnamefont {Capano}},
  \bibinfo {author} {\bibfnamefont {T.}~\bibnamefont {Dent}}, \bibinfo {author}
  {\bibfnamefont {B.}~\bibnamefont {Krishnan}}, \bibinfo {author}
  {\bibfnamefont {G.}~\bibnamefont {Meadors}}, \ and\ \bibinfo {author}
  {\bibfnamefont {A.~H.}\ \bibnamefont {Nitz}},\ }\href {\doibase
  10.1103/PhysRevD.97.124037} {\bibfield  {journal} {\bibinfo  {journal} {Phys.
  Rev. D}\ }\textbf {\bibinfo {volume} {97}},\ \bibinfo {pages} {124037}
  (\bibinfo {year} {2018})},\ \Eprint {http://arxiv.org/abs/1712.09966}
  {arXiv:1712.09966 [gr-qc]} \BibitemShut {NoStop}%
\bibitem [{\citenamefont {Testa}\ and\ \citenamefont
  {Pani}(2018)}]{Testa:2018bzd}%
  \BibitemOpen
  \bibfield  {author} {\bibinfo {author} {\bibfnamefont {A.}~\bibnamefont
  {Testa}}\ and\ \bibinfo {author} {\bibfnamefont {P.}~\bibnamefont {Pani}},\
  }\href {\doibase 10.1103/PhysRevD.98.044018} {\bibfield  {journal} {\bibinfo
  {journal} {Phys. Rev. D}\ }\textbf {\bibinfo {volume} {98}},\ \bibinfo
  {pages} {044018} (\bibinfo {year} {2018})},\ \Eprint
  {http://arxiv.org/abs/1806.04253} {arXiv:1806.04253 [gr-qc]} \BibitemShut
  {NoStop}%
\bibitem [{\citenamefont {Avishai}\ and\ \citenamefont
  {Knoll}(1976)}]{Avishai:1976bk}%
  \BibitemOpen
  \bibfield  {author} {\bibinfo {author} {\bibfnamefont {Y.}~\bibnamefont
  {Avishai}}\ and\ \bibinfo {author} {\bibfnamefont {J.}~\bibnamefont
  {Knoll}},\ }\href {\doibase 10.1007/BF01418138} {\bibfield  {journal}
  {\bibinfo  {journal} {Z. Phys. A}\ }\textbf {\bibinfo {volume} {279}},\
  \bibinfo {pages} {415} (\bibinfo {year} {1976})}\BibitemShut {NoStop}%
\bibitem [{\citenamefont {Vib\'ok}\ and\ \citenamefont
  {Balint‐Kurti}(1992)}]{Vibok:1992}%
  \BibitemOpen
  \bibfield  {author} {\bibinfo {author} {\bibfnamefont {A.}~\bibnamefont
  {Vib\'ok}}\ and\ \bibinfo {author} {\bibfnamefont {G.~G.}\ \bibnamefont
  {Balint‐Kurti}},\ }\href {\doibase 10.1063/1.462414} {\bibfield  {journal}
  {\bibinfo  {journal} {J. Chem. Phys.}\ }\textbf {\bibinfo {volume} {96}},\
  \bibinfo {pages} {7615} (\bibinfo {year} {1992})}\BibitemShut {NoStop}%
\bibitem [{\citenamefont {Berti}\ \emph {et~al.}(2009)\citenamefont {Berti},
  \citenamefont {Cardoso},\ and\ \citenamefont {Starinets}}]{Berti:2009kk}%
  \BibitemOpen
  \bibfield  {author} {\bibinfo {author} {\bibfnamefont {E.}~\bibnamefont
  {Berti}}, \bibinfo {author} {\bibfnamefont {V.}~\bibnamefont {Cardoso}}, \
  and\ \bibinfo {author} {\bibfnamefont {A.~O.}\ \bibnamefont {Starinets}},\
  }\href {\doibase 10.1088/0264-9381/26/16/163001} {\bibfield  {journal}
  {\bibinfo  {journal} {Class. Quant. Grav.}\ }\textbf {\bibinfo {volume}
  {26}},\ \bibinfo {pages} {163001} (\bibinfo {year} {2009})},\ \Eprint
  {http://arxiv.org/abs/0905.2975} {arXiv:0905.2975 [gr-qc]} \BibitemShut
  {NoStop}%
\bibitem [{\citenamefont {Poisson}\ and\ \citenamefont
  {Will}(2014)}]{Poisson:2014}%
  \BibitemOpen
  \bibfield  {author} {\bibinfo {author} {\bibfnamefont {E.}~\bibnamefont
  {Poisson}}\ and\ \bibinfo {author} {\bibfnamefont {C.~M.}\ \bibnamefont
  {Will}},\ }\href {\doibase 10.1017/CBO9781139507486} {\emph {\bibinfo {title}
  {Gravity: Newtonian, Post-Newtonian, Relativistic}}}\ (\bibinfo  {publisher}
  {Cambridge University Press},\ \bibinfo {address} {Cambridge, England},\
  \bibinfo {year} {2014})\BibitemShut {NoStop}%
\bibitem [{\citenamefont {Will}(2018)}]{Will:2018bme}%
  \BibitemOpen
  \bibfield  {author} {\bibinfo {author} {\bibfnamefont {C.~M.}\ \bibnamefont
  {Will}},\ }\href@noop {} {\emph {\bibinfo {title} {{Theory and Experiment in
  Gravitational Physics}}}}\ (\bibinfo  {publisher} {Cambridge University
  Press},\ \bibinfo {year} {2018})\BibitemShut {NoStop}%
\bibitem [{\citenamefont {Pitjeva}\ and\ \citenamefont
  {Pitjev}(2013)}]{Pitjeva:2013xxa}%
  \BibitemOpen
  \bibfield  {author} {\bibinfo {author} {\bibfnamefont {E.~V.}\ \bibnamefont
  {Pitjeva}}\ and\ \bibinfo {author} {\bibfnamefont {N.~P.}\ \bibnamefont
  {Pitjev}},\ }\href {\doibase 10.1093/mnras/stt695} {\bibfield  {journal}
  {\bibinfo  {journal} {Mon. Not. Roy. Astron. Soc.}\ }\textbf {\bibinfo
  {volume} {432}},\ \bibinfo {pages} {3431} (\bibinfo {year} {2013})},\ \Eprint
  {http://arxiv.org/abs/1306.3043} {arXiv:1306.3043 [astro-ph.EP]} \BibitemShut
  {NoStop}%
\bibitem [{\citenamefont {Gebhardt}\ \emph {et~al.}(2011)\citenamefont
  {Gebhardt}, \citenamefont {Adams}, \citenamefont {Richstone}, \citenamefont
  {Lauer}, \citenamefont {Faber}, \citenamefont {Gultekin}, \citenamefont
  {Murphy},\ and\ \citenamefont {Tremaine}}]{Gebhardt:2011yw}%
  \BibitemOpen
  \bibfield  {author} {\bibinfo {author} {\bibfnamefont {K.}~\bibnamefont
  {Gebhardt}}, \bibinfo {author} {\bibfnamefont {J.}~\bibnamefont {Adams}},
  \bibinfo {author} {\bibfnamefont {D.}~\bibnamefont {Richstone}}, \bibinfo
  {author} {\bibfnamefont {T.~R.}\ \bibnamefont {Lauer}}, \bibinfo {author}
  {\bibfnamefont {S.~M.}\ \bibnamefont {Faber}}, \bibinfo {author}
  {\bibfnamefont {K.}~\bibnamefont {Gultekin}}, \bibinfo {author}
  {\bibfnamefont {J.}~\bibnamefont {Murphy}}, \ and\ \bibinfo {author}
  {\bibfnamefont {S.}~\bibnamefont {Tremaine}},\ }\href {\doibase
  10.1088/0004-637X/729/2/119} {\bibfield  {journal} {\bibinfo  {journal}
  {Astrophys. J.}\ }\textbf {\bibinfo {volume} {729}},\ \bibinfo {pages} {119}
  (\bibinfo {year} {2011})},\ \Eprint {http://arxiv.org/abs/1101.1954}
  {arXiv:1101.1954 [astro-ph.CO]} \BibitemShut {NoStop}%
\bibitem [{\citenamefont {Do}\ \emph {et~al.}(2019)\citenamefont {Do} \emph
  {et~al.}}]{Do:2019txf}%
  \BibitemOpen
  \bibfield  {author} {\bibinfo {author} {\bibfnamefont {T.}~\bibnamefont {Do}}
  \emph {et~al.},\ }\href {\doibase 10.1126/science.aav8137} {\bibfield
  {journal} {\bibinfo  {journal} {Science}\ }\textbf {\bibinfo {volume}
  {365}},\ \bibinfo {pages} {664} (\bibinfo {year} {2019})},\ \Eprint
  {http://arxiv.org/abs/1907.10731} {arXiv:1907.10731 [astro-ph.GA]}
  \BibitemShut {NoStop}%
\bibitem [{\citenamefont {Abuter}\ \emph {et~al.}(2022)\citenamefont {Abuter}
  \emph {et~al.}}]{GRAVITY:2021xju}%
  \BibitemOpen
  \bibfield  {author} {\bibinfo {author} {\bibfnamefont {R.}~\bibnamefont
  {Abuter}} \emph {et~al.} (\bibinfo {collaboration} {GRAVITY}),\ }\href
  {\doibase 10.1051/0004-6361/202142465} {\bibfield  {journal} {\bibinfo
  {journal} {Astron. Astrophys.}\ }\textbf {\bibinfo {volume} {657}},\ \bibinfo
  {pages} {L12} (\bibinfo {year} {2022})},\ \Eprint
  {http://arxiv.org/abs/2112.07478} {arXiv:2112.07478 [astro-ph.GA]}
  \BibitemShut {NoStop}%
\end{thebibliography}%

\end{document}